\documentclass[traditabstract]{aa}

\usepackage{graphicx}
\usepackage{txfonts}
\usepackage{natbib}
\usepackage[colorlinks=true,citecolor=blue]{hyperref}
\usepackage{xspace}
\usepackage{subfig}
\usepackage{xcolor}

\newcommand{\mic}{$\mu$m\xspace}
\newcommand{\as}{\hbox{$^{\prime\prime}$}\xspace}
\newcommand{\lsd}{\hbox{$\lambda/D$}\xspace}

\begin{document}

\title{Apodization in high-contrast long-slit spectroscopy}
\subtitle{II. Concept validation and first on-sky results with VLT/SPHERE}
\titlerunning{Apodization in high-contrast long-slit spectroscopy. II.}

\author{
A. Vigan\inst{1,2} \and
M. N'Diaye\inst{3} \and
K. Dohlen\inst{1} \and 
J.-L. Beuzit\inst{4,5} \and
A. Costille\inst{1} \and
A. Caillat\inst{1} \and
A. Baruffolo\inst{6} \and
P. Blanchard\inst{1} \and
M. Carle\inst{1} \and
M. Ferrari\inst{1} \and
T. Fusco\inst{1,7} \and
L. Gluck\inst{4,5} \and
E. Hugot\inst{1} \and
M. Jaquet\inst{1} \and
M. Langlois\inst{1,8} \and
D. Le Mignant\inst{1} \and
M. Llored\inst{1} \and
F. Madec\inst{1} \and
D. Mouillet\inst{4,5} \and
A. Orign\'e\inst{1} \and
P. Puget\inst{4,5} \and
B. Salasnich\inst{6} \and
J.-F. Sauvage\inst{1,7}
}

\institute{
Aix Marseille Universit\'e, CNRS, LAM (Laboratoire d'Astrophysique de Marseille) UMR 7326, 13388, Marseille, France \\ \email{\href{mailto:arthur.vigan@lam.fr}{arthur.vigan@lam.fr}} 
\and
European Southern Observatory, Alonso de Cordova 3107, Vitacura, Santiago, Chile 
\and
Space Telescope Science Institute, 3700 San Martin Drive, Baltimore, MD 21218, USA 
\and
Universit\'e Grenoble Alpes, IPAG, F-38000 Grenoble, France  
\and
CNRS, IPAG, F-38000 Grenoble, France 
\and
INAF - Osservatorio Astronomico di Padova, Vicolo dell'Osservatorio 5, 35122 Padova, Italy 
\and
ONERA, The French Aerospace Lab BP72, 29 avenue de la Division Leclerc, 92322 Ch\^atillon Cedex, France 
\and
CRAL, UMR 5574, CNRS, Universit\'e Lyon 1, 9 avenue Charles Andr\'e, 69561 Saint Genis Laval Cedex, France 
}

\date{Received 19 October 2015; accepted 27 November 2015}
   
\abstract{Spectral characterization of young, giant exoplanets detected by direct imaging is one of the tasks of the new generation of high-contrast imagers. For this purpose, the VLT/SPHERE instrument includes a unique long-slit spectroscopy (LSS) mode coupled with Lyot coronagraphy in its infrared dual-band imager and spectrograph (IRDIS). The performance of this mode is intrinsically limited by the use of a non-optimal coronagraph, but in a previous work we demonstrated that it could be significantly improved at small inner-working angles using the stop-less Lyot coronagraph (SLLC). We now present the development, testing, and validation of the first SLLC prototype for VLT/SPHERE. Based on the transmission profile previously proposed, the prototype was manufactured using microdots technology and was installed inside the instrument in 2014. The transmission measurements agree well with the specifications, except in the very low transmissions (<5\% in amplitude). The performance of the SLLC is tested in both imaging and spectroscopy using data acquired on the internal source. In imaging, we obtain a raw contrast gain of a factor 10 at 0.3\as and 5 at 0.5\as with the SLLC. Using data acquired with a focal-plane mask, we also demonstrate that no Lyot stop is required to reach the full performance, which validates the SLLC concept. Comparison with a realistic simulation model shows that we are currently limited by the internal phase aberrations of SPHERE. In spectroscopy, we obtain a gain of $\sim$1~mag in a limited range of angular separations. Simulations show that although the main limitation comes from phase errors, the performance in the non-SLLC case is very close to the ultimate limit of the LSS mode. Finally, we obtain the very first on-sky data with the SLLC, which appear extremely promising for the future scientific exploitation of an apodized LSS mode in SPHERE.}


\maketitle
   
\section{Introduction}
\label{sec:introduction}

\defcitealias{vigan2013}{Paper I}

Detection and spectral characterization of young, giant exoplanets in the near-infrared is the primary goal of the new generation of near-infrared high-contrast imagers. These instruments have been designed to provide high-contrast at small angular separation through extreme adaptive optics (XAO) systems \citep{fusco2006,poyneer2008} and efficient coronagraphs \citep{rouan2000,soummer2005}. The back-end science subsystems of these instruments generally include spectroscopic capabilities that are made possible by diffraction-limited integral field spectroscopy \citep[IFS, e.g.,][]{antichi2009}.
This provides multi-spectral data allowing very efficient reduction of the speckle-noise through differential imaging techniques \citep{racine1999,marois2006} and enables the possibility to reach unprecedented contrasts \citep{vigan2015a}. These new instruments will hopefully provide insight into the population of giant planets, their composition, and evolution.

\begin{figure*}
  \centering
  \includegraphics[width=0.49\textwidth]{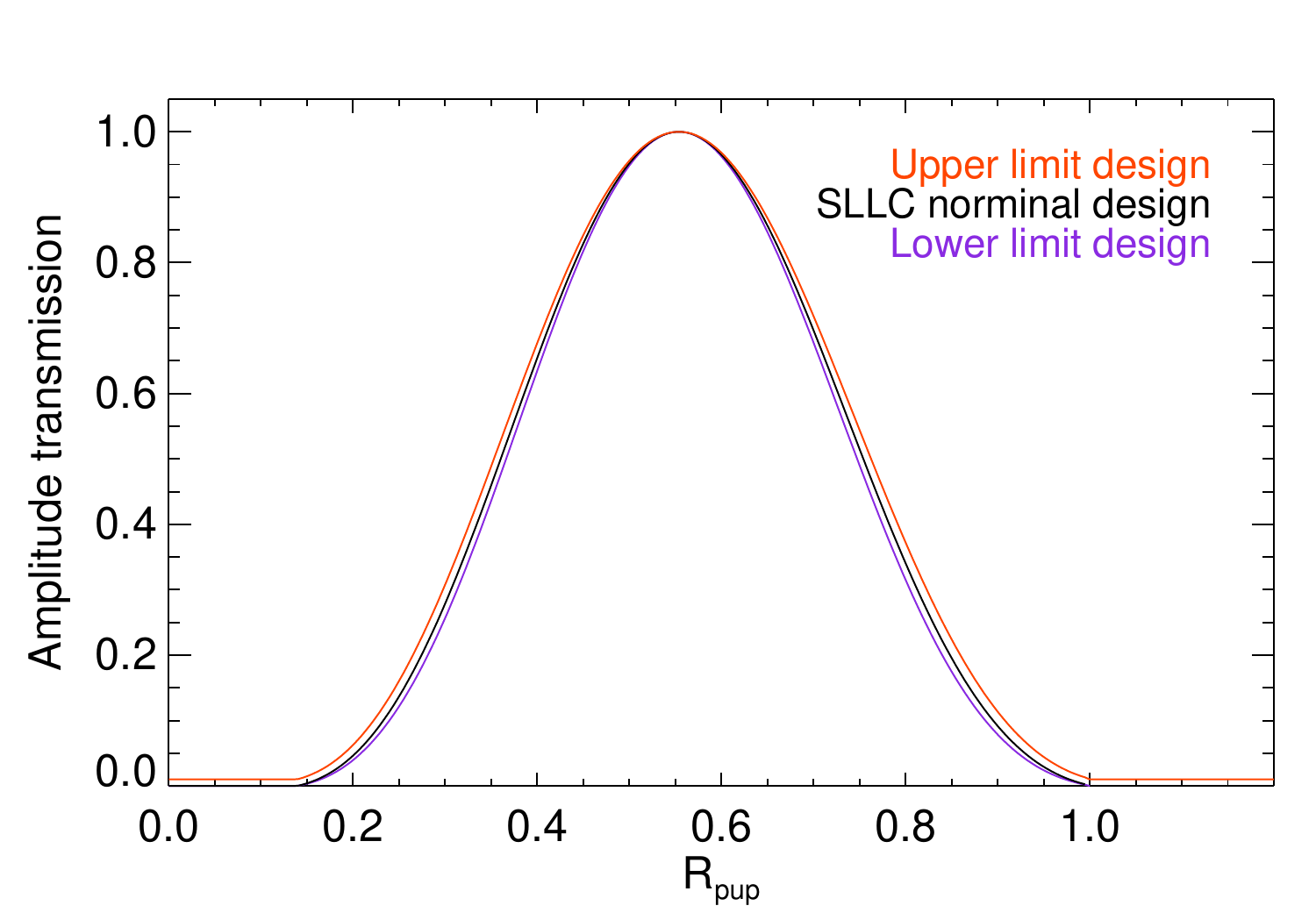}
  \includegraphics[width=0.49\textwidth]{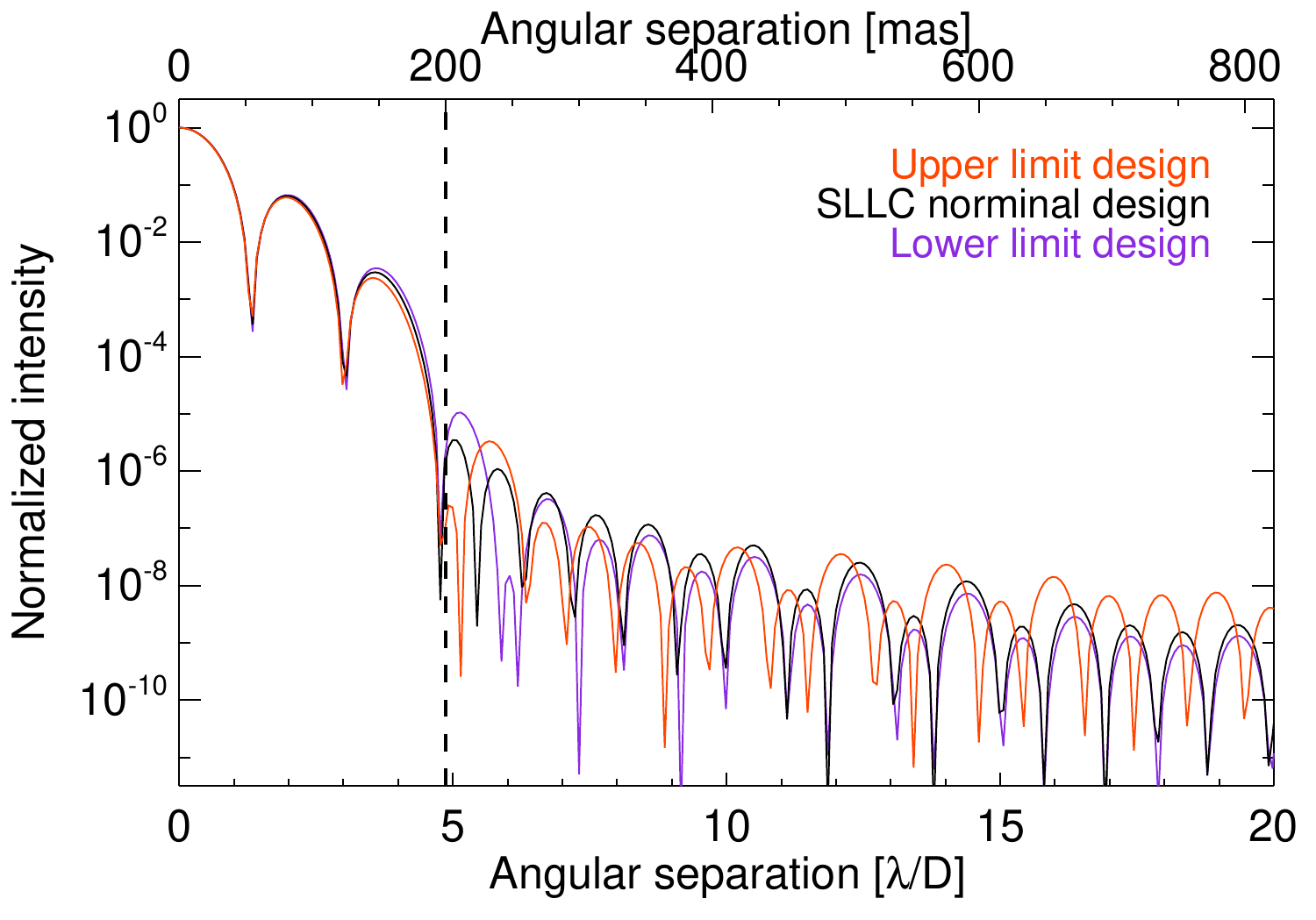}
  \caption{\emph{Left:} Amplitude transmission profile for the nominal SLLC design together with upper and lower limits for manufacturing as a function of the telescope pupil radius. \emph{Right:} Radial intensity profile of the coronagraphic images with the nominal SLLC design together with upper and lower limits, calculated at a wavelength $\lambda = 1600$~nm. The physical edge of the IRDIS coronagraphic mask in LSS mode is represented as a dashed line at 200~mas.}
  \label{fig:sllc_specs}
\end{figure*}

The Spectro-Polarimetric High-contrast Exoplanet REsearch (SPHERE) planet-finder instrument installed at the VLT \citep{beuzit2008} is a highly specialized instrument, dedicated to high-contrast imaging and spectroscopy of young giant exoplanets. It is based on the SAXO extreme adaptive optics system \citep{fusco2006,petit2014,sauvage2014}, which controls a $41\times41$ actuators deformable mirror and four control loops (fast visible tip-tilt, high-orders, near-infrared differential tip-tilt, and pupil stabilization). The common path optics employ several stress polished toric mirrors \citep{hugot2012} to transport the beam to the coronagraphs and scientific instruments. Several types of coronagraphic devices for stellar diffraction suppression are provided, including apodized pupil Lyot coronagraphs \citep{soummer2005} and achromatic four-quadrant phase masks \citep{boccaletti2008}.

In the near-infrared, SPHERE includes two scientific subsystems: an IFS \citep{claudi2008} and the infrared dual-band imager and spectrograph \citep[IRDIS;][]{dohlen2008}. The latter is a versatile instrument that includes dual-band imaging \citep{vigan2010} for the detection of exoplanets and long-slit spectroscopy (LSS) coupled with Lyot coronagraphy for their characterization \citep{vigan2008}. The LSS mode provides some unique capabilities in the high-contrast instrumentation world, the main one being that it allows reaching resolutions 5 to 10 times higher than an IFS. However, it comes with some drawbacks such as the requirement to perform field-stabilized observations, resulting in a less stable point-spread function (PSF) and a suboptimal coronagraph, where the coronagraphic mask and the slit are merged into a single device. Another  weak point
is that a well-optimized Lyot stop cannot be achieved in this mode.

In our previous work \citep[][subsequently \citetalias{vigan2013}]{vigan2013}, we proposed to use the concept of the stop-less Lyot coronagraph \citep[SLLC;][]{ndiaye2007,ndiaye2008} to improve the performance of the LSS mode. The SLLC is an apodizer optimized to essentially remove the need for any Lyot stop, at the cost of a reduced throughtput (37\% in a design for SPHERE), an enlarged inner-working angle (4.53~\lsd, with $\lambda$ the wavelength and $D$ the telescope diameter), and a decreased theoretical contrast with respect to an Apodized Pupil Lyot Coronagraph (APLC) that is optimized for imaging \citep{soummer2005}. Nonetheless, the simulations presented in \citetalias{vigan2013} promise a significant gain at small angular separations (0.2--0.5\as), making the SLLC concept an attractive solution to improve the performance of the LSS mode.

The current paper presents the design and manufacturing of the first SLLC prototype, as well as its testing and validation within the SPHERE instrument. The paper is organized as follows: in Sect.~\ref{sec:specification_manufacturing_sllc} we present the manufacturing of the optical component and the measure of its transmission; in Sect.~\ref{sec:sllc_perf_imaging} we present the experimental results obtained in imaging, and we compare them to realistic simulations of the system; in Sect.~\ref{sec:sllc_perf_spectroscopy} we present the experimental results and simulations obtained in spectroscopy; finally, we present the very first on-sky results in Sect.~\ref{sec:first_on_sky_results}, and we conclude in Sect.~\ref{sec:conclusions}.

\section{Specification and manufacturing of the SLLC}
\label{sec:specification_manufacturing_sllc}

\subsection{Specifications}
\label{sec:specifications}

The methods for calculating and optimizing the SLLC apodizer transmission profile are presented in \citetalias{vigan2013}. The nominal profile was calculated for the H-band (1600~nm) assuming a coronagraphic mask of radius 200~mas on-sky. The VLT central obscuration (14\% in pupil diameter) is taken into account in the optimization, but the spiders were not considered to avoid an asymmetric apodization shape. Indeed, such shapes cannot be considered here because the observations are performed in a field-stabilized mode to maintain the object of interest inside the slit, resulting in a rotation of the pupil during the observations. The nominal amplitude\footnote{In the remaining text we refer to amplitude transmission as the transmission of the electric field amplitude.} transmission profile and expected theoretical performance are presented in Fig.~\ref{fig:sllc_specs}.

The ultimate performance of the apodizer is driven by its transmission profile, which needs to be as close as possible to the specification. Continuous gray apodizers have traditionally been approximated by binary apodizers created using a halftone-dot process \citep{martinez2009a,martinez2009b}. The variable pixel density allows controlling the local transmission, and the transformation from a continuous two-dimension (2D) transmission map to a binarized version is performed using a simple error diffusion algorithm \citep{dorrer2007}, that is, the same type of algorithm as is used in offset printing to render grayscale levels using only black ink. As demonstrated by \citet{martinez2009a}, this technology is able to produce apodizers with the required amplitude transmission with less than 4\% of absolute error. For reference, the tolerance on the specifications for the apodizers used in the APLC of SPHERE were also below 2.5\% of error on the amplitude transmission \citep{guerri2008}.

With these constraints in mind, we investigate changes in the amplitude transmission profile and measure the impact on the expected contrast performance. Because the nominal profile is Gaussian-like, we alter the profile by multiplying it with Gaussian functions with varying standard deviation and centered on the peak of the transmission of the nominal profile ($r_{\mathrm{max}}$). The nominal profile can be either attenuated using a function of the form

\begin{equation}
  \label{eq:sllc_profile_attenuation}
  d_{\mathrm{low}} = e^{-0.5\left(\frac{r-r_{\mathrm{max}}}{\sigma}\right)^2},
\end{equation}

\noindent or amplified using a function of the form

\begin{equation}
  \label{eq:sllc_profile_amplification}
  d_{\mathrm{up}} = 2-e^{-0.5\left(\frac{r-r_{\mathrm{max}}}{\sigma}\right)^2},
\end{equation}

\noindent where $r$ is the radial position and $\sigma$ the standard deviation of the Gaussian. We vary the value of $\sigma$ to induce variations of up to 2.5\% in the absolute amplitude transmission. 

We noted in our initial simulations for \citetalias{vigan2013} the importance of the level of the transmission profile at the edge of the pupil (outer edge or central obscuration edge). The SLLC has a very strong apodization function that provides a significant attenuation of the diffraction. This, in part, comes from the fact that there is no transmission \emph{step} at the edges of the pupil, producing a very smooth transition. This condition was also taken into account when defining the tolerance on the transmission profile for manufacturing: we tested modifications of the nominal profile where the minimum transmission was offset by constant values.

In Fig.~\ref{fig:sllc_specs} we present the final tolerances that were adopted for the manufacturing, and the expected performance compared to the performance of the nominal profile. We allow a transmission of 1\% for the upper limit at the edge of the pupil. This tolerance is extended outside of the pupil and within the central obscuration to simplify manufacturing and to provide robustness against pupil misalignment. With these tolerances, we expect no measurable change in the final performance (factor <1.5 on average).

\subsection{Manufacturing of the apodizer}
\label{sec:manufacturing}

\begin{figure*}
  \centering
  \includegraphics[height=5.47cm]{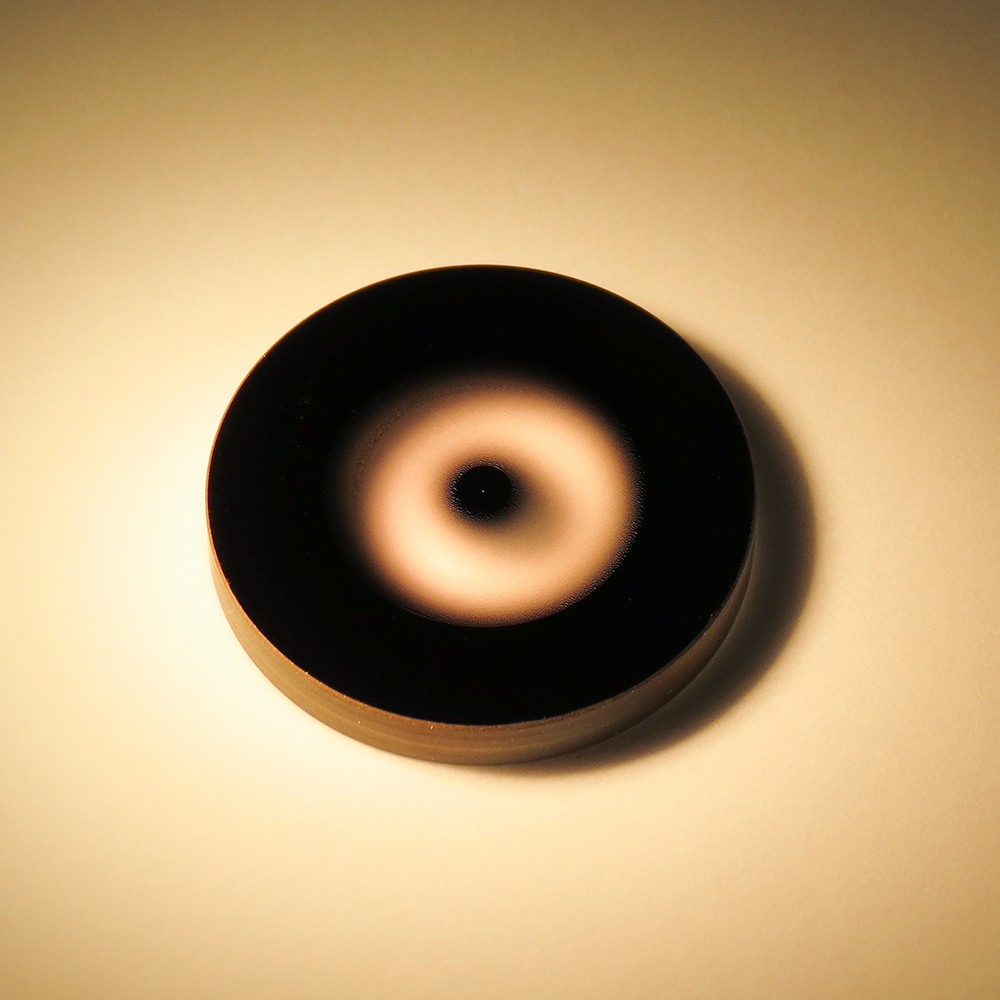}
  \hfill
  \includegraphics[height=5.47cm]{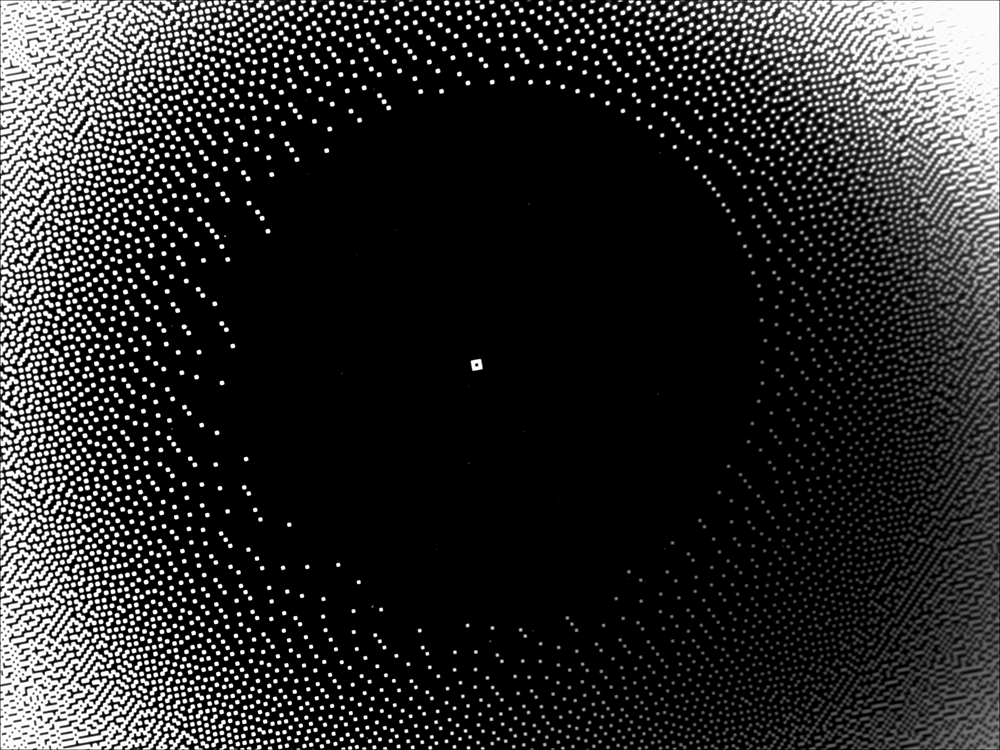}
  \hfill
  \includegraphics[height=5.47cm]{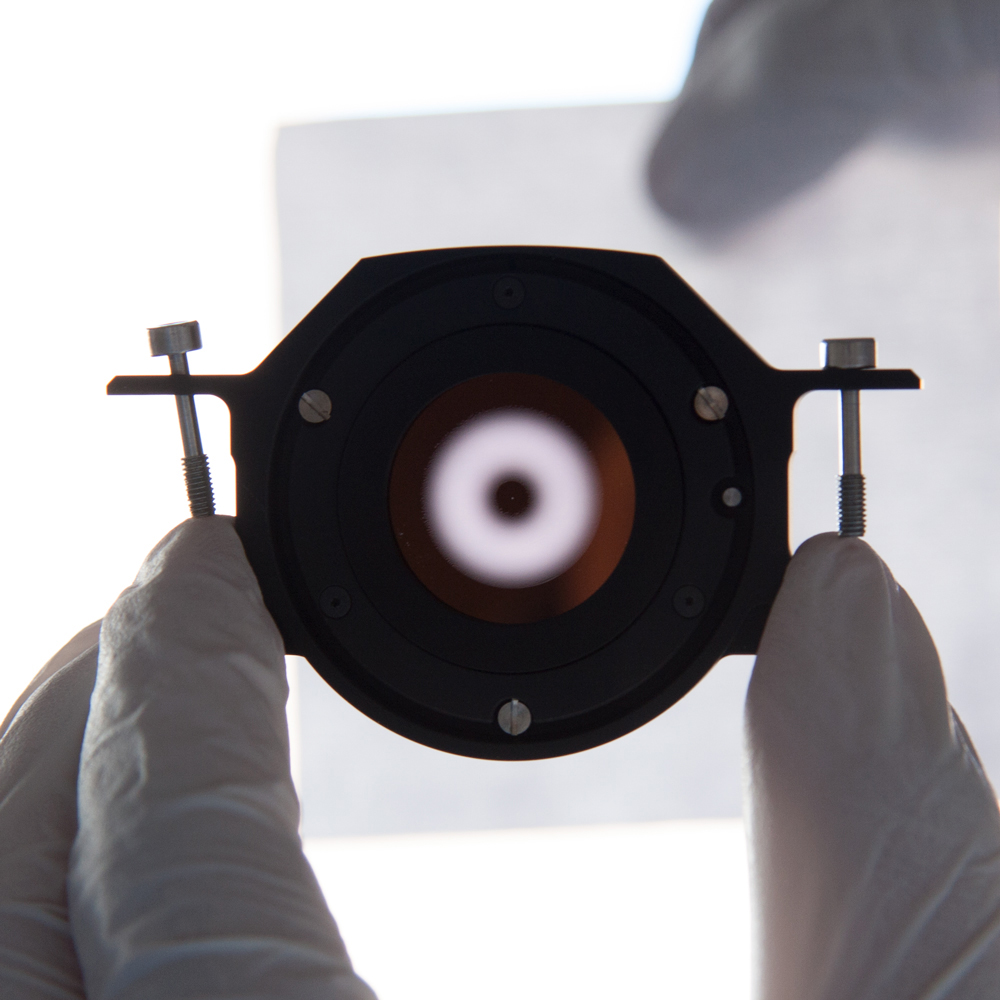}
  \caption{Chromium oxide SLLC prototype during its visual inspection in a clean room at LAM (left), under a binocular microscope (center) and in its final mechanical mount during reintegration of SPHERE in Paranal in 2014. The final SLLC component installed in SPHERE was a second prototype using an aluminum metal layer for the microdots pattern, which increases the optical density in the near-infrared compared to the chromium oxide prototype (see Sect.~\ref{sec:transmission_measurement}).}
  \label{fig:photo_sllc}
\end{figure*}

As mentioned in Sect.~\ref{sec:specifications}, gray apodizers can easily be manufactured using a microdot process (also known as halftone-dot process), where an array of ``pixels'' is deposited on a transparent substrate using lithography of a light-blocking metal layer. The density of pixels is varied to obtain the required transmission of the apodizer. A dedicated study of this technology has been performed by \citet{martinez2009a}, which demonstrated that this technology is perfectly suitable to produce the apodizers required by APLCs, without negative impact on the final coronagraphic performance. New-generation high-contrast imaging instruments SPHERE, P1640, and GPI \citep{beuzit2008,hinkley2011,macintosh2014} used APLC apodizers that were manufactured with this technology.

For the SLLC, the same technology was foreseen for the manufacturing of the prototype, but some uncertainties remained on the choice of the material to use for the metallic deposit because of the need for an extremely high optical density (>OD4 in amplitude, i.e., $10^4$ attenuation) that would allow us to achieve sufficiently close to zero transmission at the edge of the pupil. As a consequence, two SLLC prototypes were manufactured by Aktiwave LLC in Rochester, NY, USA. The substrates were made of fused-silica, on which a layer of chromium oxide was deposited for the first prototype and a layer of aluminum for the second one. The thickness of the metal layer was optimized for both metals to obtain an optical density of 4. Possible phase shift effects induced by the metal layer were not taken into account for the optimization of the pattern and its thickness. Photolithography of the metal layer was performed using a mask containing square pixels of 20~\mic in size, the density of which was optimized by Aktiwave to produce the required transmission curve. Finally, an anti-reflection coating ensuring a reflectivity R$<$1\% over the 950--2320~nm range was deposited on the other side of the substrates.

The manufactured components were received at \emph{Laboratoire d'Astrophysique de Marseille} (LAM) in 2014, where they underwent a visual inspection to check for any obvious defects. A picture of the chromium oxide prototype during its inspection is shown in Fig.~\ref{fig:photo_sllc}, with a close-up on its central part where a square pattern was introduced on purpose to allow precise centering of the apodizer in its mechanical mount. The infrared transmission profiles provided by Aktiwave agreed well with the specifications for transmissions $>$20\%, but lower transmissions could not be measured precisely.

\subsection{Transmission measurement}
\label{sec:transmission_measurement}

\begin{figure}
  \centering
  \includegraphics[width=0.5\textwidth]{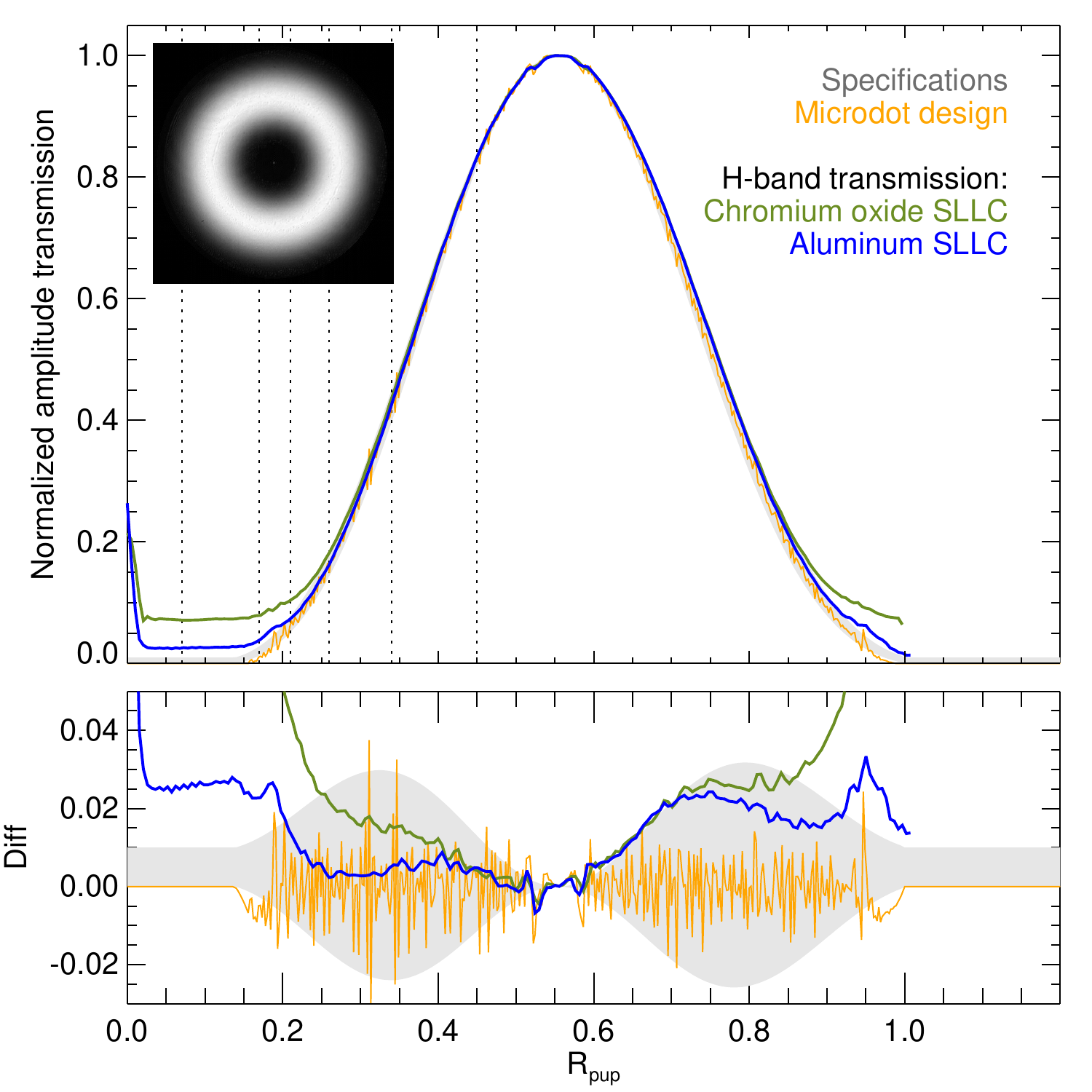}
  \caption{Azimuthal average of the amplitude transmission in $H$ band for the chromium oxide (green) and aluminum (blue) SLLC prototypes as a function of the normalized distance from the apodizer center ($\mathrm{R_{pup}}$). The increase in transmission close to zero corresponds to the pattern introduced in the design to facilitate the opto-mechanical centering of the apodizer. The design specifications are represented by a gray envelope. The transmission of the microdots map designed by Aktiwave is also plotted in orange. This profile appears noisier because the microdots map is binarized. The top plot shows the normalized transmission measurement, and the bottom plot shows the error with respect to the nominal transmission profile. The vertical dotted lines correspond to the radii where we plot the transmission as a function of wavelength in the left panel of Fig.~\ref{fig:sllc_transmission_wavelength}. The transmission map of the aluminum prototype in $H$ band is shown as an inset in the top left corner of the plot.}
  \label{fig:sllc_transmission}
\end{figure}

\begin{figure*}
  \centering
  \includegraphics[width=0.48\textwidth]{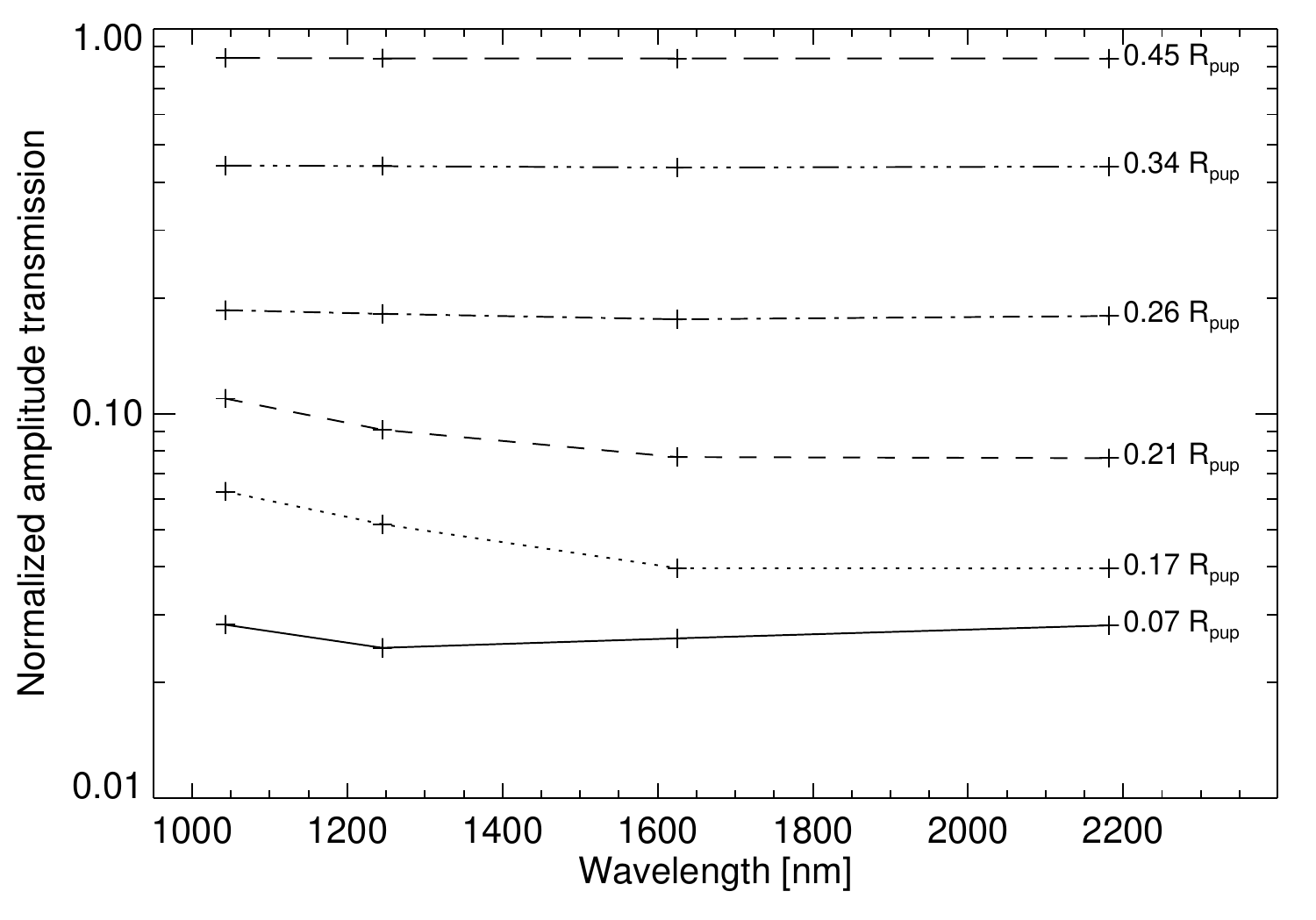}
  \includegraphics[width=0.48\textwidth]{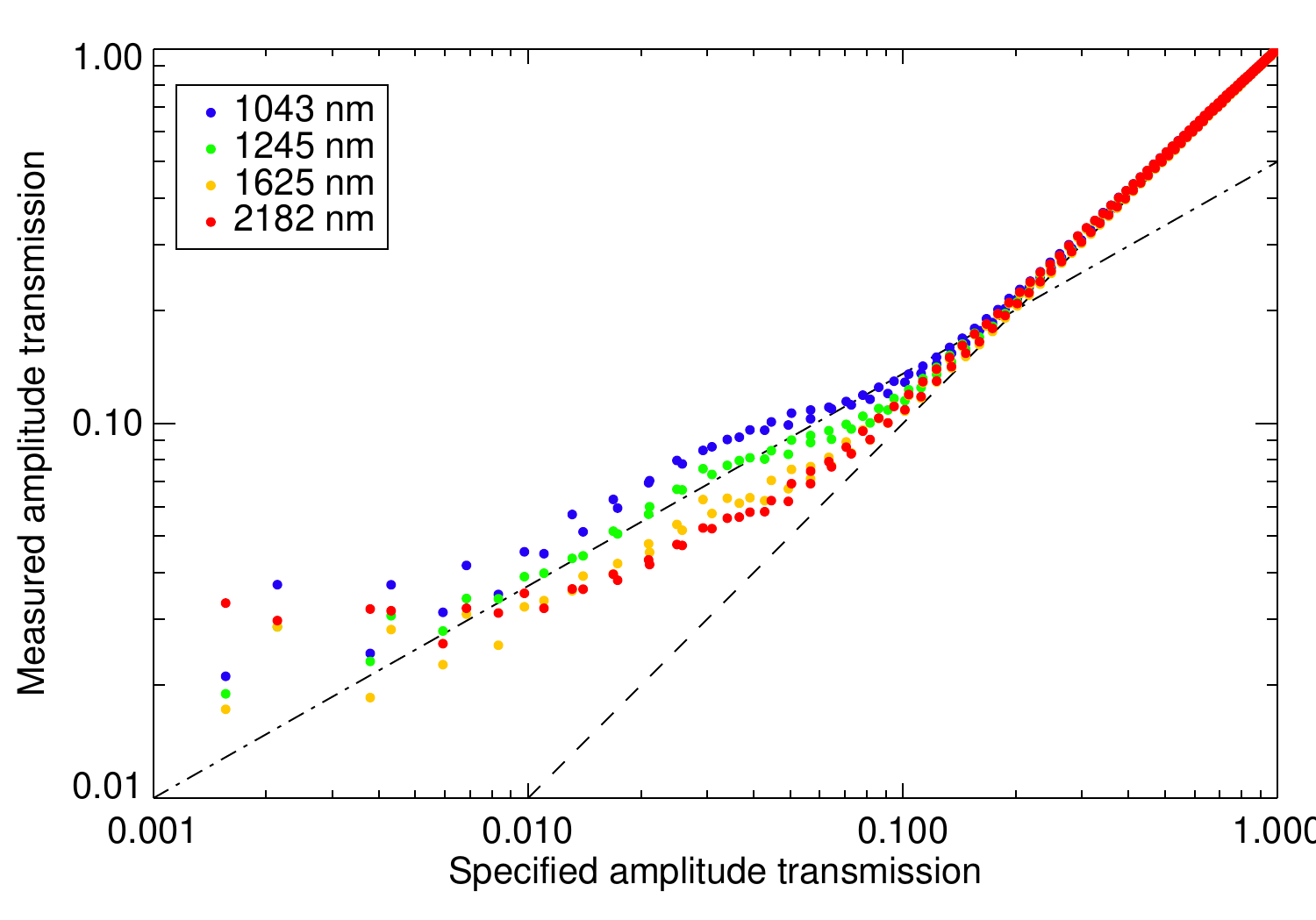}
  \caption{Chromaticity of the amplitude transmission of the aluminum prototype. \emph{Left:} Normalized amplitude transmission as a function of wavelength at different radii in the pupil (dotted vertical lines in Fig.~\ref{fig:sllc_transmission}), for which the value is indicated on the right side of each curve. \emph{Right:} Measured amplitude transmission at the four wavelengths as a function of the original specification. The dashed and dash-dotted lines show the 1 decade/decade and 0.5 decade/decade trends.}
  \label{fig:sllc_transmission_wavelength}
\end{figure*}

The SLLC prototypes were installed and quickly tested inside SPHERE during the reintegration of the instrument at the Paranal observatory in early April 2014. The transmission of both prototypes was measured by acquiring pupil images using the IRDIS pupil imaging mode. For each transmission measurement, three images are acquired: (1) a clear pupil image without the SLLC, (2) an apodized pupil image with the SLLC, and (3) a background frame with the light source switched off. Particular care was taken to adjust the integration time so as to avoid any saturation in the images. Images were acquired at four different wavelengths, with broad-band $Y$ ($\lambda = 1043$~nm, $\Delta\lambda = 70$~nm), $J$ ($\lambda = 1245$~nm, $\Delta\lambda = 120$~nm), $H$ ($\lambda = 1625$~nm, $\Delta\lambda = 145$~nm), and $K_s$ ($\lambda = 2180$~nm, $\Delta\lambda = 150$~nm) filters. The clear and apodized pupil images were background subtracted, and the bad pixels were corrected using a sigma-clipping procedure. Then, the apodized pupil image was divided by the clear pupil image to obtain the transmitted intensity. Finally, mean azimuthal profiles were calculated starting from the center of the pupil. Amplitude transmission measurements in $H-$band for the two prototypes are presented in Fig.~\ref{fig:sllc_transmission}.

From the very first measurements it appeared that the chromium oxide prototype did not meet the specifications in the low-transmission part (<20\%) of the apodizer. Because only one free position was available in the SPHERE apodizer wheel and because limited time was available for extensive tests during the reintegration period, we chose to keep the aluminum prototype
inside the instrument, which is closer to the specifications. For this reason, we do not discuss the results from the chromium oxide prototype in more detail, and we refer to \citet{vigan2014} for some additional information and measurements. From now on, we refer only to the aluminum prototype.

For the aluminum prototype, the transmission is closer to the specifications, but there are still some problems in the low-transmission parts of the apodizer. The minimum transmission obtained in the central part and at the edge of the pupil is $\sim$2.5\%. This measurement illustrates the difficulty of manufacturing apodizers for the near-IR with very high optical densities, as was already identified by previous studies in the context of the GPI instrument \citep{sivaramakrishnan2009}.

A microdots apodizer relies on the absorption and diffraction of light by small (20~\mic in our case) metallic dots deposited on a substrate whose distribution is such that the density of dots produces the required transmission profile. As detailed by \citet{dorrer2007}, the spatial filtering produced by the following optical train, that is, by the field-stop located in the coronagraph mask plane in SPHERE, causes the intensity distribution in the relayed image of the apodizer to be the square of its near-field transmission function, regardless of the wavelength. The specified near-field transmission should therefore be equal to the desired amplitude transmission function.

The measurements made on the SLLC prototype in SPHERE indicate that this theory breaks down for low-transmission values, where the expected transmission is no longer achieved (Fig.~\ref{fig:sllc_transmission}) and a clear chromatic behavior is observed (Fig.~\ref{fig:sllc_transmission_wavelength}). An indication of what happens can be gained from the right panel of Fig.~\ref{fig:sllc_transmission_wavelength}, where we plot
the measured amplitude transmission versus the specified amplitude transmission, in which we identify three distinct transmission regions:

\begin{enumerate}
  \item Amplitude transmission above around 15\%: The slope equals 1~decade/decade for all wavelengths, hence fully validating the \citet{dorrer2007} theory and producing a gray apodization function as expected.
  \item Amplitude transmission below 15\%, down to around 1\%: The transmission is higher than expected and has a notable blue tint. The slope of the curves tend toward 0.5~decade/decade. In this range, the absorbing dots agglomerate into fully opaque patches, leaving only isolated dot-sized holes. Spatial filtering therefore operates on the transmitted and not on the absorbed light, so that the measured intensity is now proportional to the near-field apodizer transmission. The effect of this is to make the measured amplitude transmission equal to the square root of the specified electric field transmission, as observed.
  \item Amplitude transmission below 1\%: Here, the apodizer is essentially fully covered with the aluminum film, and we measure the transmission of the materiel itself. The measured intensity transmission saturates at around $10^{-3}$, representing an optical density of 3, and has a slightly red tint.
\end{enumerate}

These measurements are particularly interesting for the future development of apodizer prototypes that require very low transmissions. In particular, they highlight the fact that the production of a transmission map based on a simple error-diffusion algorithm may not always be appropriate, and that further considerations are required to produce a transmission profile within tight specifications.

For the parts where the transmission is >15\%, the lower panel of Fig.~\ref{fig:sllc_transmission} shows that the profile is mostly within specifications, but we can identify two distinct regimes as a function of the pupil radius: (1) for $R_{pup} < 0.56$, which corresponds to the peak of the transmission, the profile is close to the expected transmission given by the microdots map, and (2) for $R_{pup} > 0.56$ the profile is just within specification, but it departs from the profile given by the microdots map. The most likely explanation is that the physical size of the apodizer does not exactly match that of the pupil in the apodizer plane (18~mm in diameter from design). The diameter of the pupil has been measured to be 385.2~pixel in the IRDIS focal plane. We cannot physically measure the diameter of the pupil in the apodizer plane, but we note that if we modify the measured value by only 1 pixel (386.2~pixel, i.e., an increase of 0.26\%), the measured azimuthal average falls well within the specification for $R_{pup} > 0.56$. While a pupil diameter of 386.2~pixel is not compatible with our measurements of the pupil size on the IRDIS detector, it can be translated into the apodizer plane. In this plane, a difference of only 47~\mic (0.26\% of 18~mm) between the physical pupil diameter and the specification is enough to produce the change. Such a small error on the size of the beam is plausible in the system and agrees with the tolerances.

A final important measurement is the total intensity transmission of the prototype, which we measure to be $22.9\% \pm 0.7\%$ using the four wavelengths. This is a significant attenuation that is directly related to the very strong apodization function of the SLLC, but also to a peak amplitude transmission (at $r_{\mathrm{max}}$) of only 83\% instead of >95\% as originally specified. The origin of this discrepancy has not been identified by the manufacturer.

\section{SLLC performance in imaging}
\label{sec:sllc_perf_imaging}

\subsection{Data acquisition and processing}
\label{sec:img_data_acquisition}

\begin{table*}
  \caption{Data acquisition log}
  \label{tab:data_acquisition_log}
  \centering
  \begin{tabular}{cccccccl}
  \hline\hline
  Light source & Neutral density & Apodizer & Coronagraph       & Lyot stop         & Filter      & T$_{\mathrm{exp}}$ & Comment \\
               &                 &          &                   &                   &             & (min)              &         \\
  \hline
  \multicolumn{8}{c}{Imaging and saturated imaging data} \\
  \hline
  Point source & 3.5             & No       & No                    & No                     & H2          & 3.0      & Reference PSF      \\
  Point source & 1.0             & No       & No                    & No                     & H2          & 3.0      & Saturated PSF      \\
  Point source & $\ldots$        & No       & No                    & No                     & H2          & 3.0      & Background         \\
  \hline
  Point source & 3.5             & SLLC     & No                    & No                     & H2          & 8.0      & Reference SLLC PSF \\
  Point source & 1.0             & SLLC     & No                    & No                     & H2          & 8.0      & Saturated SLLC PSF \\
  Point source & 1.0             & SLLC     & CLC                   & No                     & H2          & 8.0      & SLLC + FPM         \\
  Point source & $\ldots$        & SLLC     & No                    & No                     & H2          & 8.0      & Background         \\
  \hline
  Flat field   & $\ldots$        & No       & No                    & No                     & H2          & $\ldots$ & Flat field         \\
  \hline
  \multicolumn{8}{c}{Long-slit spectroscopy data} \\
  \hline
  Point source & No              & No       & Slit\tablefootmark{a} & Prism\tablefootmark{b} & YJHKs       & 11.0     & Coronagraphic data       \\
  No           & $\ldots$        & No       & Slit\tablefootmark{a} & Prism\tablefootmark{b} & YJHKs       & 11.0     & Background               \\
  Point source & 3.5             & No       & Slit\tablefootmark{a} & Prism\tablefootmark{b} & YJHKs       &  4.0     & Reference (off-axis) PSF \\
  No           & $\ldots$        & No       & Slit\tablefootmark{a} & Prism\tablefootmark{b} & YJHKs       &  4.0     & Background               \\
  \hline
  Point source & No              & SLLC     & Slit\tablefootmark{a} & Prism\tablefootmark{b} & YJHKs       & 12.0     & Coronagraphic SLLC data  \\
  No           & $\ldots$        & SLLC     & Slit\tablefootmark{a} & Prism\tablefootmark{b} & YJHKs       & 12.0     & Background               \\
  Point source & 3.5             & SLLC     & Slit\tablefootmark{a} & Prism\tablefootmark{b} & YJHKs       &  8.0     & Reference SLLC PSF       \\
  No           & $\ldots$        & SLLC     & Slit\tablefootmark{a} & Prism\tablefootmark{b} & YJHKs       &  8.0     & Background               \\
  \hline
  Flat field   & $\ldots$        & No       & No                    & No                     & YJHKs       & $\ldots$ & Flat field              \\
  Laser lines  & $\ldots$        & No       & Slit\tablefootmark{a} & Prism\tablefootmark{b} & YJHKs       & $\ldots$ & Wavelength calibration  \\
  \hline
  \multicolumn{8}{c}{ZELDA measurements} \\
  \hline
  Point source & No              & No       & ZELDA                 & No                     & Fe~{\sc ii} & 0.5      & ZELDA measurement     \\
  Point source & No              & No       & No                    & No                     & Fe~{\sc ii} & 0.5      & Reference clear pupil \\
  No           & $\ldots$        & No       & No                    & No                     & Fe~{\sc ii} & 0.5      & Background            \\
  \hline
  \end{tabular}
  \tablefoot{\tablefoottext{a}{The slit has a width of 0.12\as on sky and includes a central focal plane mask of radius 0.20\as.} \tablefoottext{b}{The dispersive element in IRDIS is located immediately behind the Lyot stop plane. It includes a circular Lyot stop with a size of 92\% of the pupil diameter.}}
\end{table*}

The SLLC apodizer was first tested with SPHERE/IRDIS in imaging to understand its performance and limitations. The data were acquired with SPHERE on August 31, 2015 during daytime technical time. All measurements were made internally using the light sources available in the calibration unit of the instrument \citep{wildi2009}. The acquired data are listed in Table~\ref{tab:data_acquisition_log}. For imaging, the data consisted of two sets with and without the apodizer. In each set, a deep reference PSF in the H2 filter ($\lambda = 1593$~nm, $\Delta\lambda = 26$~nm) of IRDIS was acquired with a neutral density (ND) filter of value 3.5, which provides an attenuation of a factor 955 of the flux in this filter. Then a saturated PSF was acquired by changing to a ND of 1.0 (attenuation of a factor 7.2 in filter H2). Finally, for imaging with the SLLC, an additional image was acquired with a circular, suspended focal-plane mask (FPM) of diameter 0.45\as (but no Lyot stop) to validate the SLLC concept. Corresponding calibrations were also acquired: background frames with the same integration time as the science data, and a flat-field to correct for inter-pixel sensitivity variations.

The data were all processed in a similar fashion. Master backgrounds, bad-pixel maps and flat-fields were created using the v0.15.0 release of the SPHERE data reduction and handling software \citep{pavlov2008}. Each image was first background subtracted and then divided by the flat field. Bad pixels identified in the flat and backgrounds were corrected by replacing them with the median of neighboring good pixels. Then the images were normalized by their integration time, and the attenuation of the ND was also compensated for to be able to compare the images taken with different ND filters. The saturated and unsaturated images were aligned together manually using the Airy rings and speckles, providing an accuracy of $\sim$0.1~pixel. The SLLC image with a FPM was also aligned with respect to the saturated SLLC image. Finally, azimuthal average profiles were calculated on each of the images.

\subsection{Experimental results}
\label{sec:img_experimental_results}

\begin{figure*}
  \centering
  \includegraphics[width=0.48\textwidth]{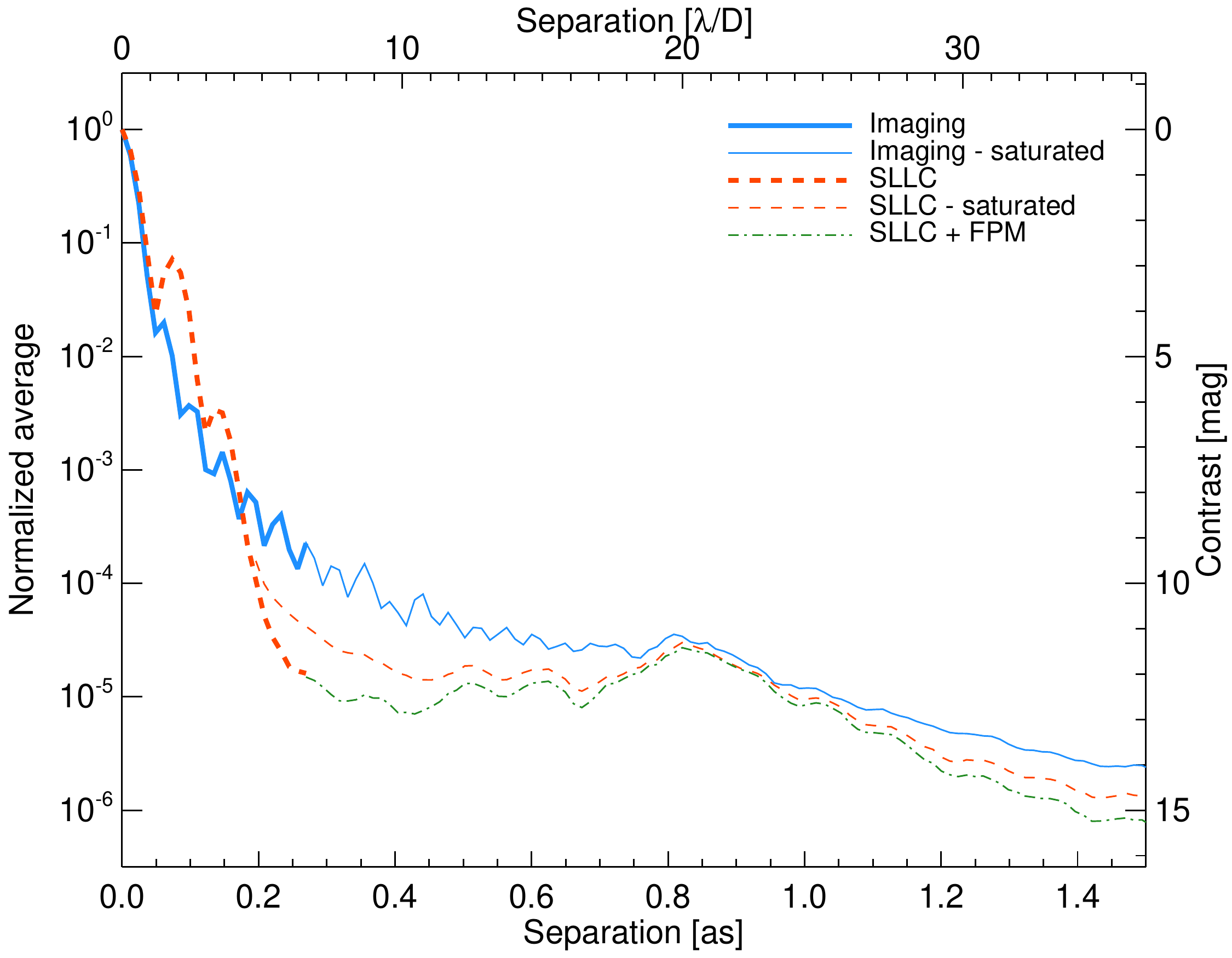}
  \hfill
  \includegraphics[width=0.48\textwidth]{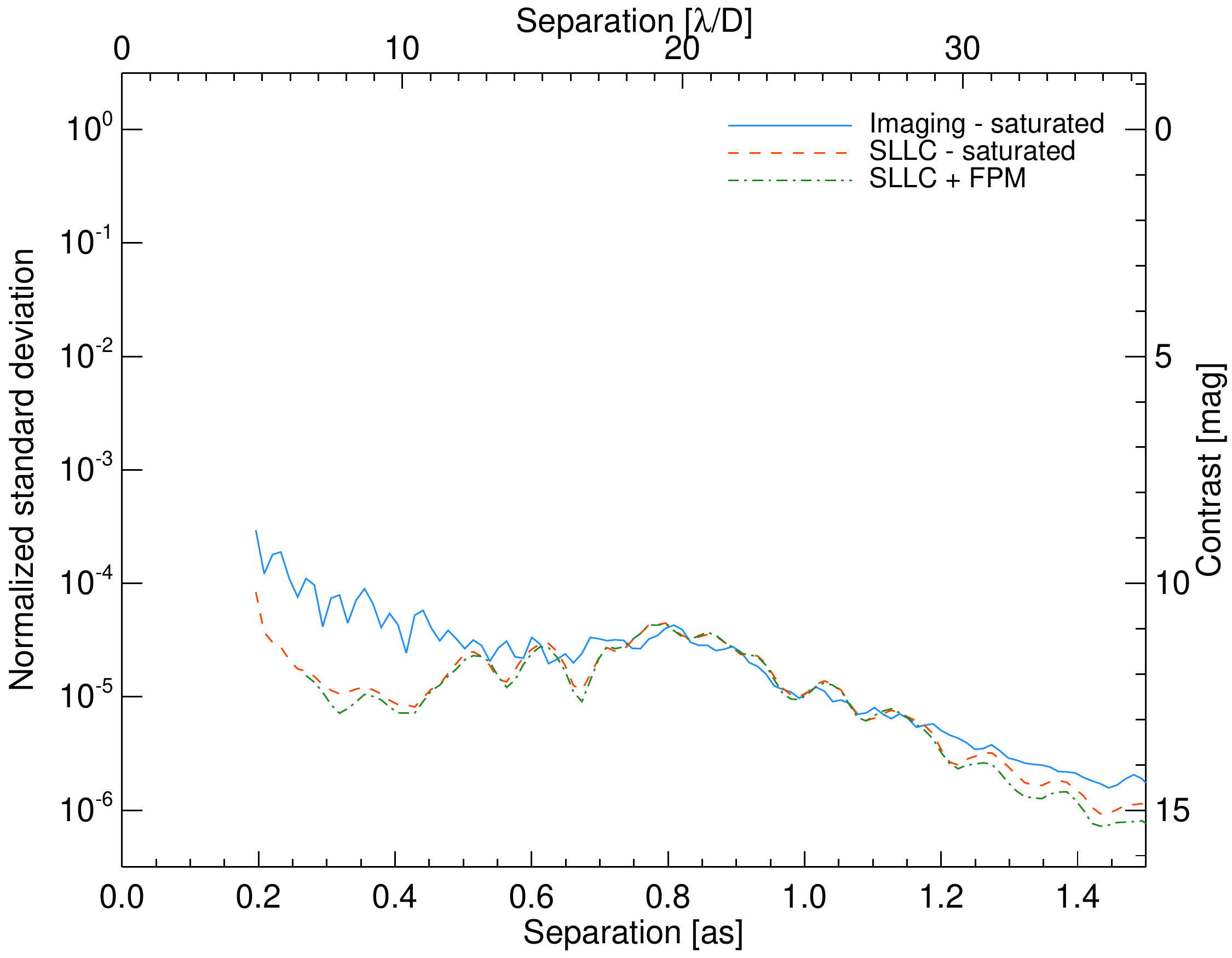}
  \caption{Normalized azimuthal average (left) and standard deviation (right) profiles as a function of angular separation for the imaging data (blue), the SLLC data (red), and the SLLC+FPM data (green) at 1593~nm. The saturated and unsaturated data overlap in the 0.20--0.25\as range. The data with the FPM plane mask is also plotted starting at 0.25\as. For the standard deviation, only the values measured on the saturated data are plotted.}
  \label{fig:saturated_imaging_data}
\end{figure*}

The experimental results obtained in imaging are presented in Fig.~\ref{fig:saturated_imaging_data}. The left panel shows the azimuthal average profile for the imaging data, the SLLC data and the SLLC+FPM data, normalized in each case to the maximum of the unsaturated PSF acquired in the same instrumental configuration. The saturated and unsaturated data overlap in the 0.20--0.25\as range, and the SLLC+FPM data are plotted starting at 0.25\as.

For the unsaturated data, we see the change of shape of the SLLC PSF, with the characteristic two bright Airy rings that were visible in Fig.~\ref{fig:sllc_specs}, followed by a steep drop of the PSF profile down to slightly over $10^{-5}$ at 0.25\as ($\sim$6~\lsd). This is different from the imaging data, for which the PSF follows a classical Airy pattern, reaching slightly over $10^{-4}$ at 0.25\as. At this radius we thus obtain a factor $\sim$10 in raw contrast. 

The saturated data are less straightforward to interpret from the azimuthal average of the SLLC data. While the saturated and unsaturated imaging data match exactly, the level of the SLLC saturated data is slightly higher in the 0.20--0.25\as range than the unsaturated SLLC data. The origin of this difference is obvious when looking at the data: because the PSF core is highly saturated, there are (1) electronic ghosts clearly visible in the images\footnote{Repetition of the saturated pattern every four readout strips on the detector (4$\times$32=128 pixels).}, (2) stray light close to the optical axis and in the AO-corrected area, and (3) a visible ghost reflection of apodizer in the upper part of the image. These different effects all contribute to affecting the final contrast in the saturated SLLC data. The last two contributions, which affect the data mostly close to the center, are at low spatial frequencies. To verify that this is an artificial effect, we plot in the right panel of Fig.~\ref{fig:saturated_imaging_data} the normalized azimuthal standard deviation of the saturated data. In this plot, the level of the SLLC profile is lower than in the azimuthal average plot, reaching the exact same level as the unsaturated PSF in the 0.20--0.25\as range. This is a good indication that the spurious contributions listed previously are only low-spatial frequencies that would not affect the detectability of point-like sources in the data. This plot shows that the SLLC delivers a gain by a factor 10 at 0.3\as and 5 at 0.5\as, which is on the order of our predictions in \citetalias{vigan2013} in the presence of a realistic amount of aberrations (>50~nm~rms).

Finally, we also compared the saturated data with the data acquired with SLLC+FPM. In this data set, we did not use a Lyot stop, that is, we only used  the FPM as an anti-saturation device. The azimuthal standard deviation curve exactly matches the one from saturated imaging, which means that we obtain an identical performance in the two configurations. In addition, we note that for the SLLC+FPM data set, the azimuthal average and azimuthal standard deviation profiles match almost exactly, which demonstrates that we are truly in the static speckle noise regime \citep{goodman1975,soummer2007}. This is also a validation of the SLLC concept: it shows that with this device, we do not need to use a Lyot stop to remove diffraction, contrary to the classical Lyot coronagraph (CLC) or APLC, for which both a FPM and a Lyot stop are necessary to achieve complete diffraction suppression.

\subsection{Comparison to simulations}
\label{sec:img:comparison_simulations}

\begin{figure}
  \centering
  \includegraphics[width=0.5\textwidth]{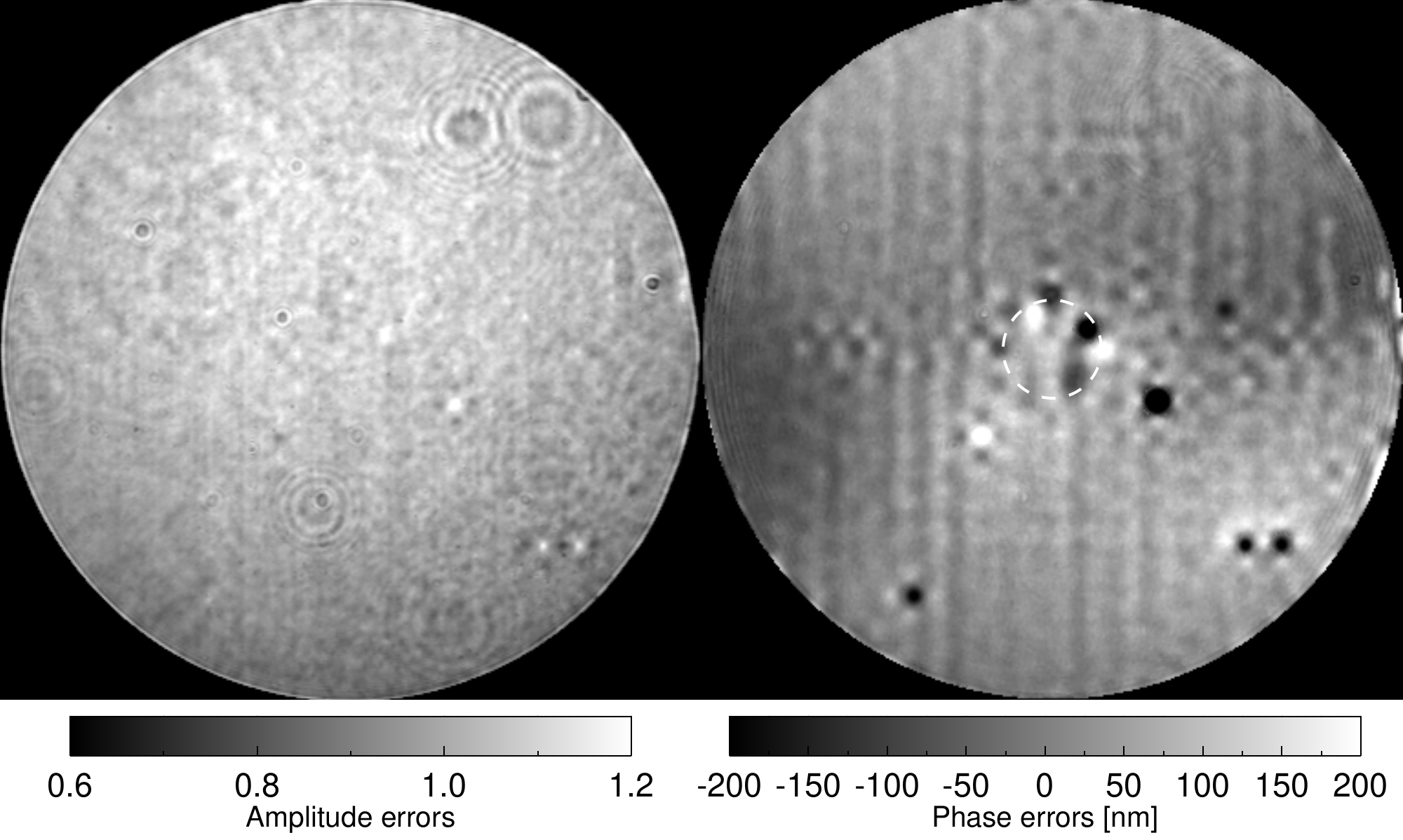}
  \caption{Two inputs used in our SPHERE simulation model. Left, a map of the amplitude errors measured in the broad-band $H$ filter and normalized to a median value of 1.0. Right, a map of the phase errors measured with the prototype ZELDA wavefront sensor \citep{ndiaye2013} installed in SPHERE (see text for details). In this map, the dead or stuck actuators of the SPHERE DM are easily visible as white or black circular spots. The visible actuators at the edge of the central obscuration (dashed circle, 14\% of the pupil in diameter) are not dead or stuck, but they are not controlled properly due to their significant overlap with the central obscuration. Without taking into account the dead or stuck actuators, there is $\sim$35~nm~rms of aberrations inside the system.}
  \label{fig:simulation_inputs}
\end{figure}

\begin{figure}
  \centering
  \includegraphics[width=0.49\textwidth]{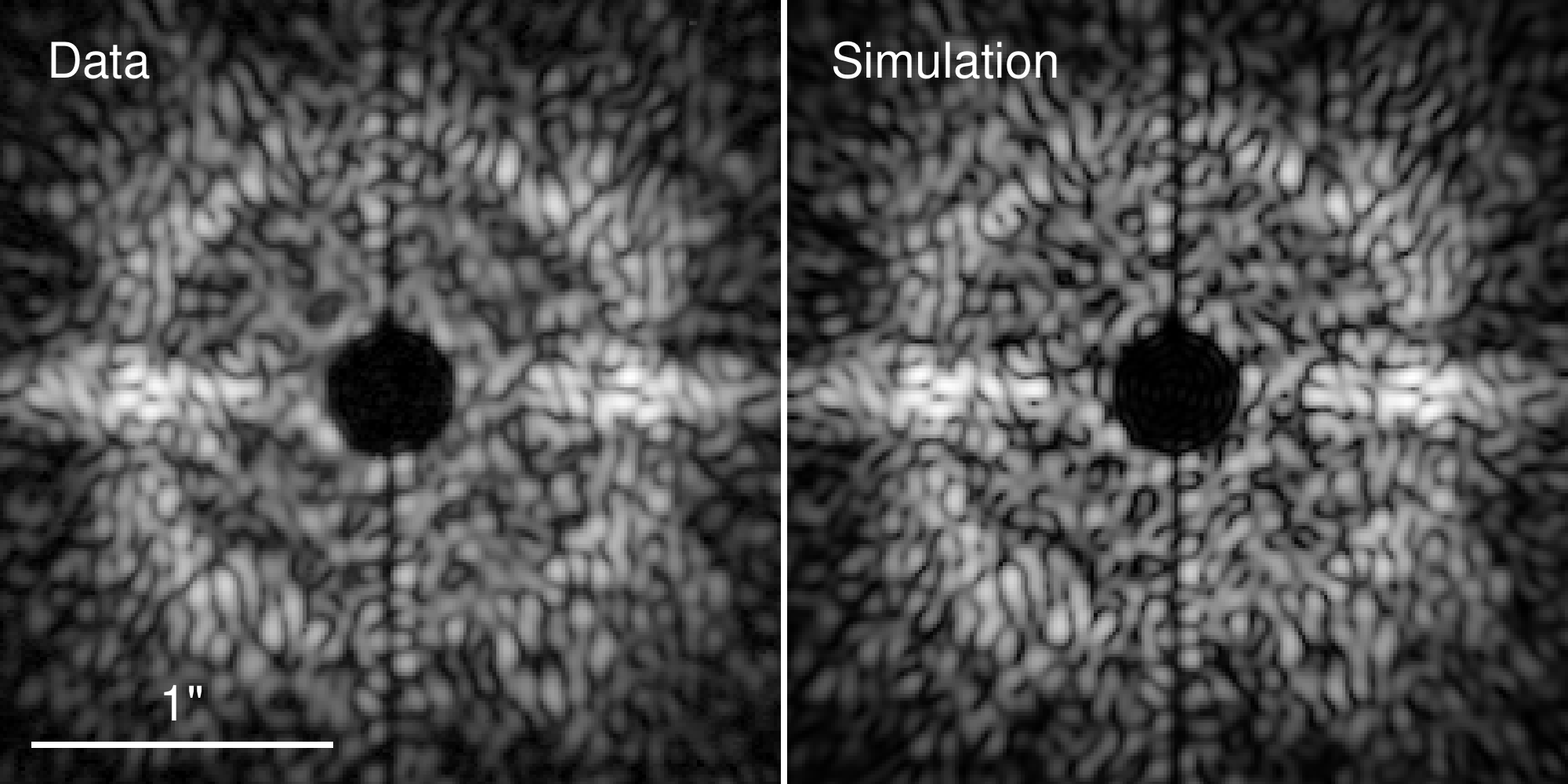}
  \caption{Comparison of SLLC+FPM imaging data acquired in the IRDIS H2 filter (left) with the output of our SPHERE simulation model (right). The vertical dark line at the center corresponds to the arms that hold the suspended FPM. The spatial extension is 2.6\as on the side.}
  \label{fig:data_simu_images}
\end{figure}

\begin{figure}
  \centering
  \includegraphics[width=0.48\textwidth]{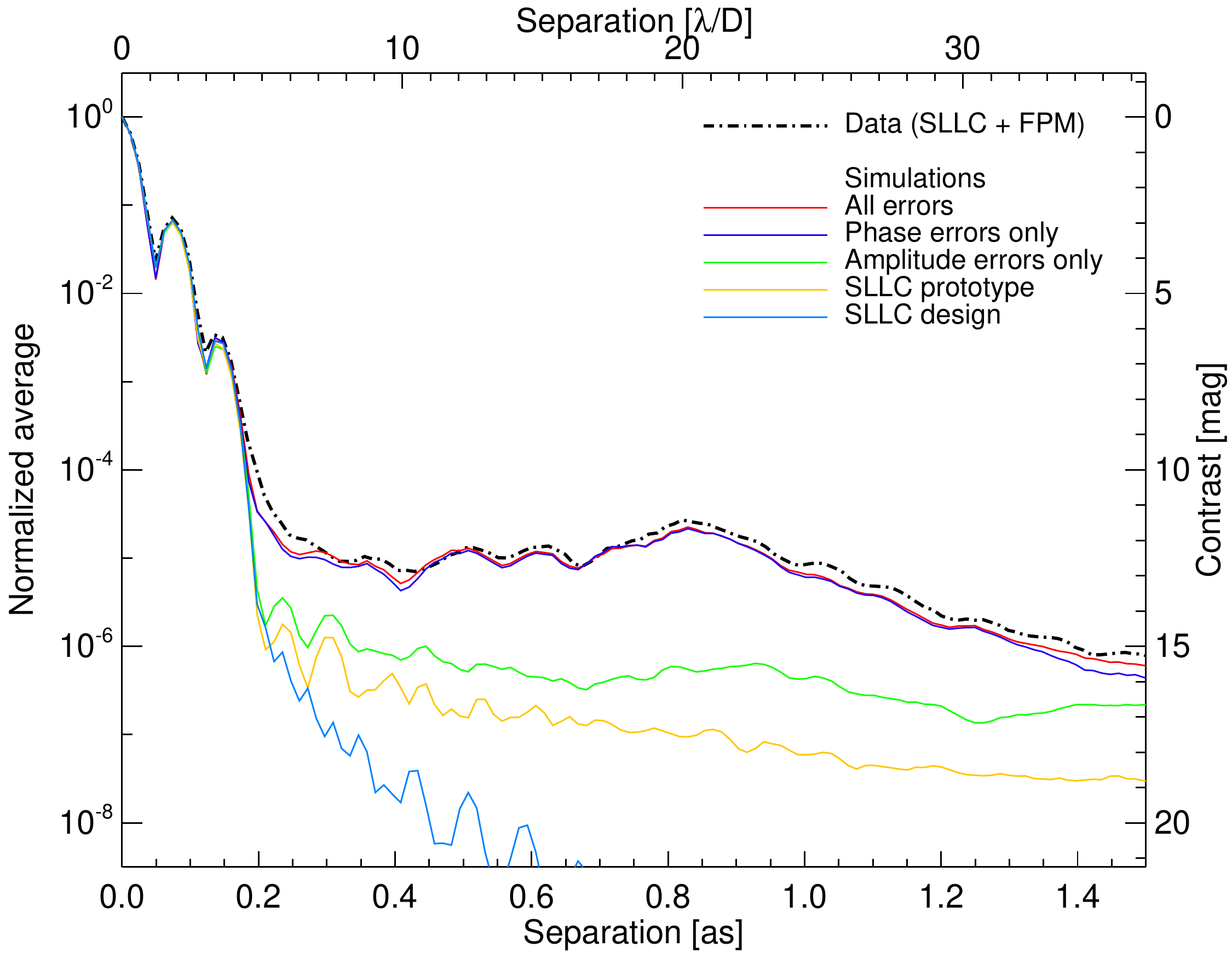}
  \caption{Normalized azimuthal average for different simulations in the H2 filter (1593~nm) with the SLLC and taking into account various error terms: all sources of errors (red), only phase errors (dark blue), only amplitude errors (green), only the measured transmission map of the SLLC prototype (orange), and finally with only the SLLC nominal design (light blue). The SLLC+FPM data are overplotted as a black dash-dotted line.}
  \label{fig:saturated_imaging_simu}
\end{figure}

To better understand our data and the limitations of the system, we built a simple but realistic simulation model of SPHERE/IRDIS. We first define the complex amplitude of the electric field with phase and amplitude errors measured in the instrument (see below). We then propagate the obtained electric field through our coronagraph in four successive steps that are described as follows:
\begin{enumerate}
    \item The electric field is apodized by the SLLC in the entrance pupil plane using the transmission map measured in Sect.~\ref{sec:transmission_measurement}.
    \item With a Fourier transform of the pupil plane field, we derive the electric field in the following focal plane and filter out the resulting field with an opaque mask (FPM) to block the central part of the source light.
    \item By means of an inverse Fourier transform of the focal plane field, we produce the relayed pupil plane field over a stop whose size corresponds to the diameter of the optics in SPHERE. As a reminder, the SLLC does not need to make use of a Lyot stop, but the residual light in this Lyot plane is still blocked by the optics outer part. In IRDIS, the optics are only oversized by 10\% with respect to the beam size in the Lyot pupil plane.
    \item With a Fourier transform of the re-imaged pupil plane field, we finally obtain the electric field in the final image plane. A squared modulus of the resulting electric field provides us with the image produced by the SLLC in the presence of wavefront errors.
\end{enumerate}

The electric fields were computed over the area of interest in each plane of the system using the semi-analytical method with Matrix Fourier transform as suggested by \citet{soummer2007} for a fast and efficient computation of Lyot-style coronagraphs. The final simulated images were modeled to observe the Nyquist-Shannon sampling theorem at a wavelength of 950~nm to exactly match the spatial scale of the IRDIS data. This resulted in a scale of 12.25~mas/pixel in the image plane.

For the injection of the amplitude errors in our simulation model, we used an image of the pupil of the instrument measured in the relayed pupil plane in the absence of coronagraph and simply used its square root value to obtain the amplitude (Fig.~\ref{fig:simulation_inputs}, left). In this image some of the dead or stuck actuators are
clearly evident, as well as what appears to be specks of dust in out-of-pupil surfaces. The overall pupil is also slightly non-uniformly illuminated and appears to be brighter on the upper left part than in the lower right part. This effect is due to a very small angular misalignment of the internal fiber light source and is not visible on-sky. 

To accurately estimate the phase errors, we used ZELDA, a Zernike phase mask sensor that was also installed in SPHERE during the reintegration of the instrument to measure the quasi-static coronagraphic aberrations in the instrument \citep{ndiaye2013,ndiaye2014,dohlen2013}. This sensor uses a phase mask centered on the stellar signal in the coronagraphic focal plane to code the phase wavefront errors in the entrance pupil plane into intensity variations in the relayed pupil plane. Our phase mask prototype introduces a phase shift of 0.222\,$\lambda_0$ over a 1.087~$\lambda_0$/D diameter at the wavelength $\lambda_0=1626$~nm. We operated the sensor on the internal source with the Fe~{\sc ii} near-infrared narrow-band filter centered at $\lambda_0$ (see Table~\ref{tab:data_acquisition_log}). The dynamic range of ZELDA is derived based on these characteristics and following \citet{ndiaye2013}. In the present case, phase errors between -0.136 and 0.364~$\lambda_0$ (-223~nm and 598~nm) on the wavefront can theoretically be measured without ambiguity.

As SPHERE was specified to produce a maximum of 36~nm~rms internal wavefront error, our measurements are theoretically performed in the small aberration regime. Based on this assumption, we work in practice around the null phase functioning point and derive wavefront aberrations in the range $\pm$100~nm with sub-nanometric accuracy from the relayed pupil intensity data, using a phase reconstruction with a second-order polynomial expression. Since the SPHERE deformable mirror (DM) of the instrument presents several dead or stuck actuators in the pupil (Sauvage et al., submitted), phase errors at these locations are so large that they go beyond the dynamical range of the Zernike sensor, generating phase wrapping effects in the intensity measurements, and therefore making the estimate of these errors extremely tedious and inaccurate to extract. To obtain realistic phase error values at the location of these actuators, we replaced the actuator response points by values that follow a Gaussian function. For each actuator, we chose the characteristics of this function, relying on the points outside the actuator and on the phase wrapping observed in the ZELDA data. Obviously, this only constitutes a rough estimate, which proved enough for us, however, to obtain a first model of our experimental SLLC data. The right panel of Fig.~\ref{fig:simulation_inputs} shows the phase error map derived from ZELDA estimate and modeling of the dead actuators response. A 35~nm rms wavefront error is estimated with ZELDA over the pupil after masking these actuators, showing excellent consistency with the value predicted from the system analysis study of SPHERE during the instrument phase A (36~nm rms; \citealt{boccaletti2008}; \citealt{dohlen2011}).

Using this SPHERE model, we simulated data in the IRDIS H2 filter. A visual comparison between the SLLC+FPM data and the corresponding simulation is shown in Fig.~\ref{fig:data_simu_images}, and a normalized azimuthal average profile that compares the data with simulations including different error terms is presented in Fig.~\ref{fig:saturated_imaging_simu}. The visual agreement in Fig.~\ref{fig:data_simu_images} is excellent: all the main structures are visible, and many common speckles or groups of speckles can be identified between the data and simulation. The match is not perfect, however, certainly because the inputs of our model are not perfect measurements. In particular, as mentioned above, the phase map measured with ZELDA does not exactly represent the aberrations within the system because of the dead or stuck actuators. In addition, our model does not include either chromatic effects or Fresnel propagation effects. Even though these effects are expected to be small in SPHERE \citep{dohlen2011}, they might contribute slightly at the level of individual speckles. Finally, ZELDA only measures the aberrations upstream of the coronagraph, which means that the aberrations introduced downstream are not taken into account. For IRDIS, these aberrations amount to a maximum of $\sim$21~nm~rms (see error budget in \citealt{dohlen2008}).

Figure~\ref{fig:saturated_imaging_simu} confirms that the model is nonetheless very good. In this figure, the normalized average profile measured on the data is compared to simulations including various error terms. There is an excellent match between the data and the profile that includes all error terms (phase and amplitude errors, and the measured SLLC transmission map), which proves that our SPHERE simulation model is enough to understand the current limitations of the SLLC in SPHERE. Clearly, the highest contribution comes from the phase errors in the range 4--30~\lsd, with the amplitude errors being a factor 5 to 20 lower, depending on the angular separation. At separations >35~\lsd, the amplitude errors start to be noticeable, but hopefully they are mostly static and their contribution can be removed through differential imaging. The contribution of the transmission of the SLLC prototype is mostly negligible above contrast ratios of $10^{-6}$, but we see a significant departure between the performance with the measured profile and with the nominal design. 

In conclusion, the SLLC essentially allows us to reach the current limit of the system in imaging, which is defined by the level of phase errors ($\sim$35~nm~rms). However, the SLLC was initially designed for the LSS mode of IRDIS, therefore we now analyze the performance of this mode with the SLLC in more detail.

\section{SLLC performance in spectroscopy}
\label{sec:sllc_perf_spectroscopy}

\subsection{Data acquisition and processing}
\label{sec:spectro_data_acquisition}

The LSS data were acquired on the same day and in the same conditions as the imaging data presented in Sect.~\ref{sec:img_data_acquisition}. We acquired two data sets in the low-resolution spectroscopy (LRS) mode of IRDIS, one with the SLLC and one without (Table~\ref{tab:data_acquisition_log}). In this mode, a slit is located in the coronagraphic focal plane. It has a width of 0.12\as and includes a central FPM of radius 0.20\as. The following Lyot pupil plane includes a slightly undersized circular Lyot stop with a size equal to 92\% of the pupil diameter. The dispersive element, a double prism, is located immediately after this basic Lyot stop. The double prism and Lyot stop constitute a single opto-mechanical component, and they cannot be decoupled. As a result, we are effectively working with a circular Lyot stop in LSS. However, as demonstrated in \citetalias{vigan2013}, this stop is not optimized and does not provide efficient diffraction suppression at small separations, which is what triggered the development of the SLLC for the IRDIS/LSS mode.

The centering of the PSF on the FPM at the center of the slit was performed manually so as to optimize it as best possible. Then the differential tip-tilt loop of SPHERE ensures that this centering remains constant during all subsequent exposures in closed loop \citep[see, e.g.,][]{vigan2015a}. For each data set, we acquired a deep coronagrapic image with the PSF behind the FPM, and an off-axis reference PSF where the PSF is offset inside the slit with respect to the FPM. Necessary calibrations were also acquired: background frames with the same integration time as the science data, a flat-field to correct for inter-pixel sensitivity variations, and a wavelength calibration.

The data were treated in a way very similar to the imaging data, with first a subtraction of the background, then a division by the flat-field, and finally a correction of the bad pixels. The SPHERE pipeline was also used to reduce the wavelength calibration, which allowed us to attribute the corresponding wavelength to each pixel. All spectra were normalized by their integration time, and for the off-axis reference PSF, the effect of the ND filter that was used to avoid saturation was compensated for at each wavelength. 

Finally, coronagraphic profiles were extracted at different wavelengths along the spatial dimension and were normalized to the peak of the off-axis reference PSF at the same wavelength. Because LSS provides only a single spatial dimension and a single spectral dimension (contrary to an IFS, for example), it is not possible to calculate an azimuthal average profile. However, to take into account the finite width of the slit (0.12\as, or $\sim$10 pixels on the detector), we averaged the profile along the spectral dimension over a width of 10~pixels to account for all the flux at the considered wavelength.

\subsection{Experimental results and comparison to simulations}
\label{sec:spectro_results}

\begin{figure*}
  \centering
  \includegraphics[width=1.0\textwidth]{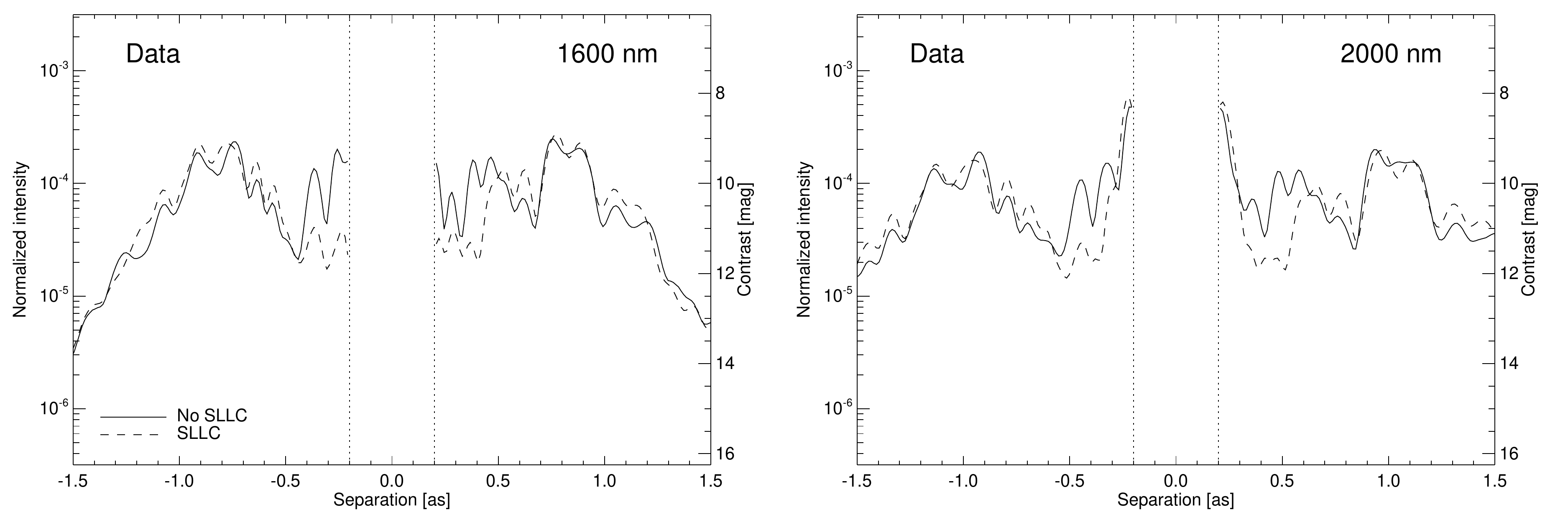}
  \includegraphics[width=1.0\textwidth]{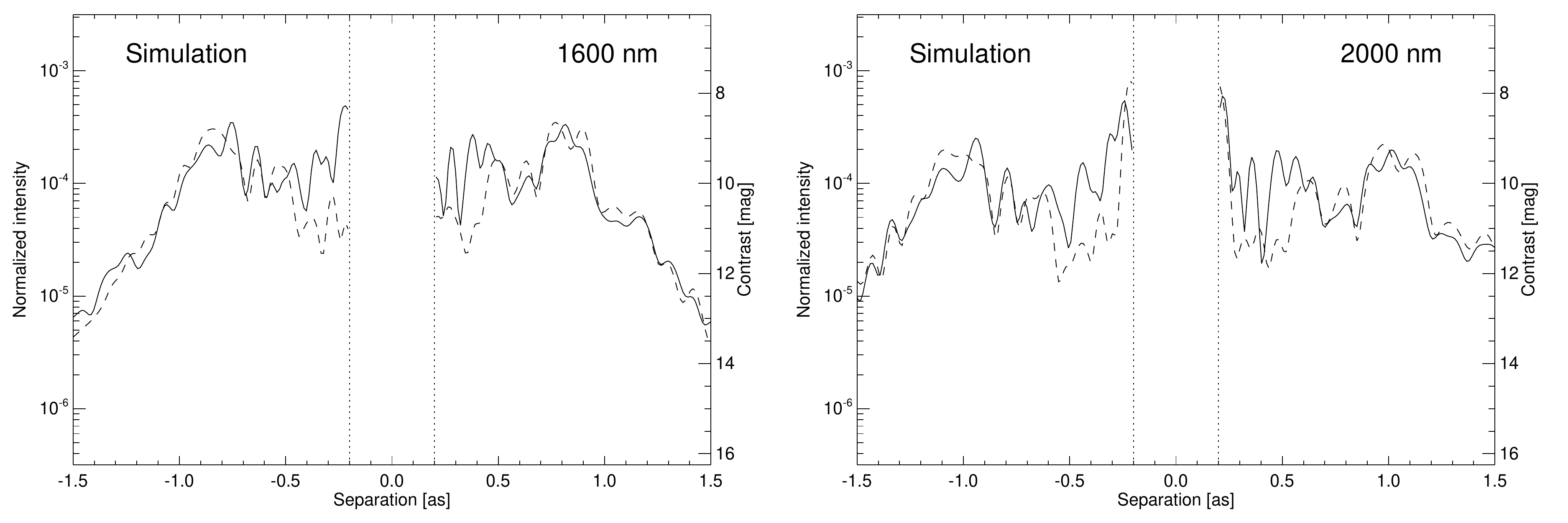}
  \caption{IRDIS/LSS coronagraphic intensity profiles with (dashed line) and without (plain line) the SLLC as a function of angular separation on both sides of the optical axis and normalized with respect to an off-axis PSF. Profiles are plotted at two wavelengths, 1600~nm and 2000~nm. The top panel presents the data acquired in SPHERE and the bottom panel the simulations performed at the same wavelengths with our SPHERE simulation model including all error terms. The central part between -0.2\as and 0.2\as without
any data corresponds to the location of the FPM inside the slit.}
  \label{fig:spectro_data_simu_comparison}
\end{figure*}

\begin{figure}
  \centering
  \includegraphics[width=0.5\textwidth]{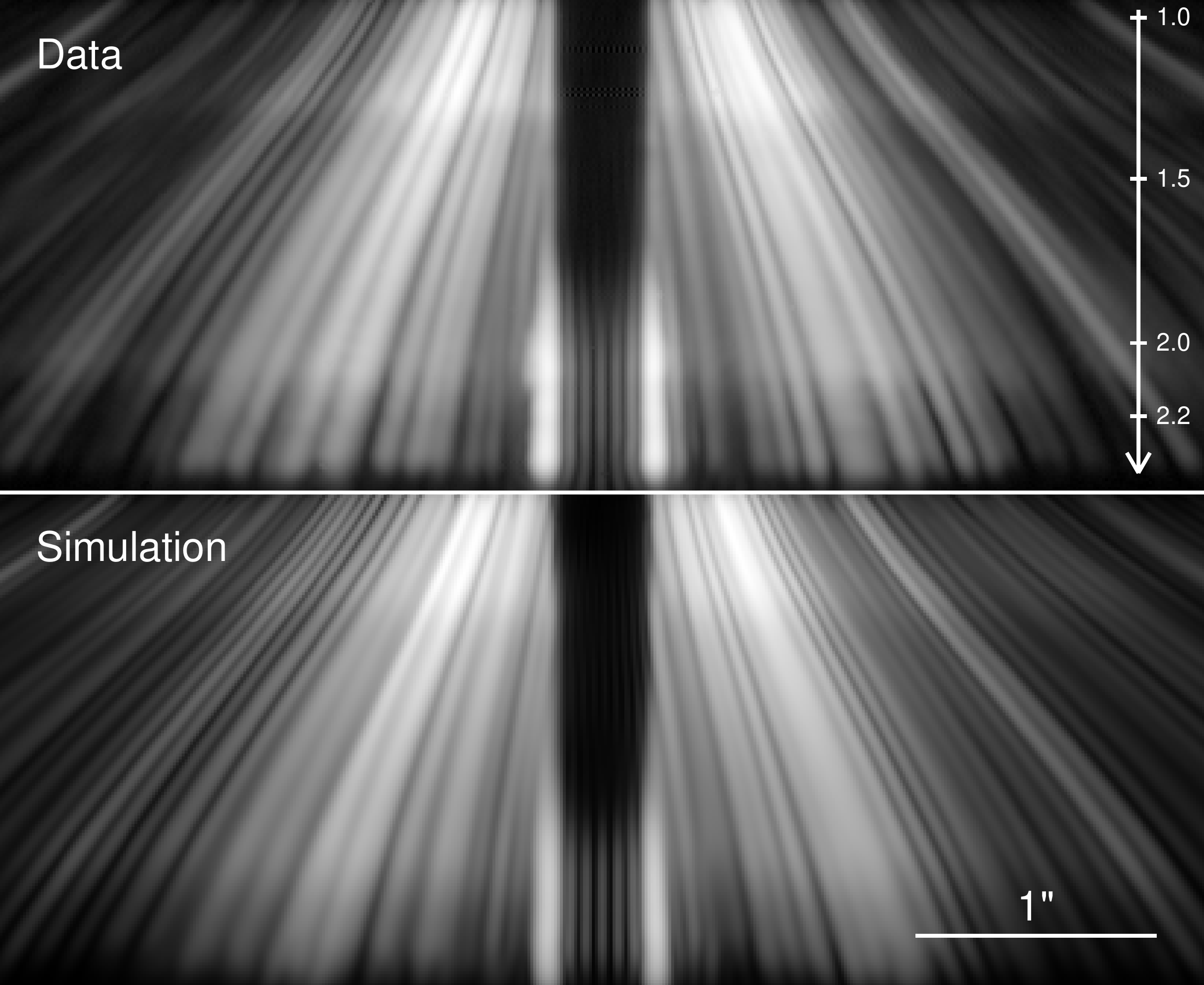}
  \caption{IRDIS/LSS data acquired with the SLLC (top) compared to the output of our SPHERE simulation model (bottom). The dark band at the center corresponds to the location of the FPM inside the slit. The wavelength scale (in micron) is indicated with the downward white arrow. The total spectral range is 0.95--2.3~\mic, and the spatial extension is 2.5\as.}
  \label{fig:data_simu_spectra}
\end{figure}

\begin{figure*}
  \centering
  \includegraphics[width=1.0\textwidth]{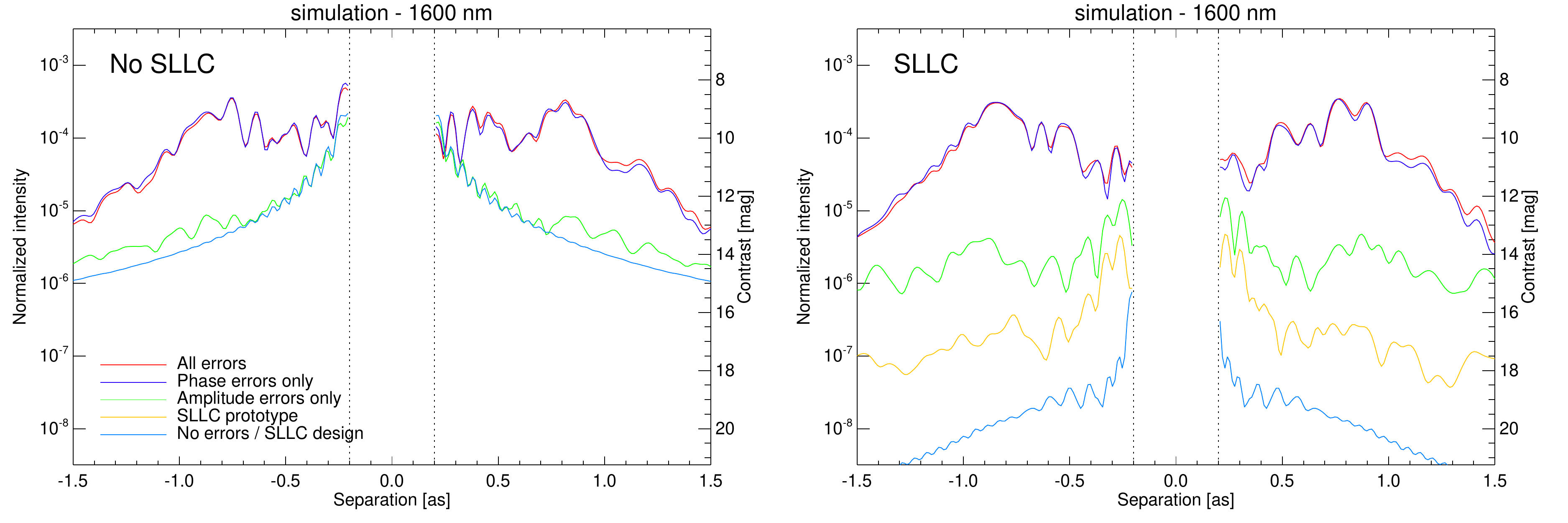}
  \caption{IRDIS/LSS coronagraphic intensity profiles simulated with (right) and without (left) the SLLC at 1600~nm and taking into account various error terms: all sources of errors (red), only phase errors (dark blue), only amplitude errors (green), only the measured transmission map of the SLLC prototype (orange), and finally with only the SLLC nominal design (light blue).}
  \label{fig:spectro_simu}
\end{figure*}

The results with and without SLLC are compared at two different wavelengths, 1600~nm and 2000~nm, in the top panel of Fig.~\ref{fig:spectro_data_simu_comparison}. The wavelength at 1600~nm corresponds to the wavelength at which the SLLC was optimized given the FPM width of 0.2\as, while that
at 2000~nm corresponds to a longer wavelength at which the mode is not as well optimized given the size of the FPM. At both wavelengths there is a range of angular separations where the SLLC provides a clear gain ($\sim$1~mag) with respect to the case without SLLC. At 1600~nm, this range is 0.2--0.5\as, which corresponds approximately to the range where some gain was expected in \citetalias{vigan2013}. Beyond this range, both profiles are at a very similar level. At 2000~nm, the first Airy ring of the PSF in the SLLC case starts to be larger than the FPM, and there is a clear leakage on both sides of the FPM. This leakage is again present in the non-apodized case because the FPM is also too small for this mode in $K$-band. However, in the range between 0.3\as and 0.6\as  the SLLC still measurably improves the result by $\sim$1~mag.

We also compared the data to simulations produced with our model of the instrument. The LSS data were simulated by introducing a slit with FPM in the coronagraphic focal plane and a circular Lyot stop with a size equal to 92\% of the pupil diameter in the relayed pupil plane. Focal-plane images of the slit were simulated at the same wavelengths as the data ($\sim$170 independent wavelengths), dispersed according to the same wavelength law, and co-added to form a spectrum. A visual comparison between the SLLC data and the simulation is presented in Fig.~\ref{fig:data_simu_spectra}, and coronagraphic profiles at 1600~nm and 2000~nm are presented in the bottom panel of Fig.~\ref{fig:spectro_data_simu_comparison}. We see again that the data and the simulation match well. The orders of magnitude are the same, and the range of separations where the SLLC brings a gain of $\sim$1~mag is similar. This good match makes us again confident that our simulation model is robust and that the observed gain with the SLLC is also real in the LSS mode.

However, the match is not as perfect as for the imaging case in Fig.~\ref{fig:saturated_imaging_simu}. The reason is that in LSS we only have access to a single spatial dimension, so we cannot calculate an azimuthal average that would smooth the local differences and only show the underlying speckle field statistics. Indeed, the slit samples only the speckles that are visible along the y-axis in Fig.~\ref{fig:data_simu_images} (the slit is horizontal in SPHERE) and disperses them, so that any difference in the speckle field at that location will be immediately visible in the coronagraphic profile. We have seen in Sect.~\ref{sec:img:comparison_simulations} that the speckles are not all exactly reproduced by our simulation, which means that some differences between the data and the simulation are to be expected in LSS. This effect is visible in Fig.~\ref{fig:data_simu_spectra}, but the very good match for the imaging data confirms that our model is valid and robust.

In Fig.~\ref{fig:spectro_simu} we use our simulation model to estimate the level of the different error contributions and see how it compares with and without SLLC. The most striking conclusion is that despite the fact that the performance is currently limited by the phase errors (as for the imaging case, see Sect.~\ref{sec:img:comparison_simulations}), the ultimate limit of the non-SLLC case is on average a factor 100 lower than the ultimate limit with the SLLC\footnote{This is when considering the transmission of the manufactured SLLC prototype measured in Sect.~\ref{sec:transmission_measurement}. With the nominal transmission, the factor is close to 1000.}. This confirms that the current LSS design is a poor coronagraph. In particular, very close to the edge of the FPM (0.2--0.4\as), the performance without SLLC appears already close to the ultimate limit (factor <2), while for the SLLC the level of the amplitude errors is still a factor $\sim$5 below, and the ultimate limit is a factor 10--20 below. This means that the non-apodized LSS mode will very quickly reach its ultimate limit if the performance of the system can be improved through the use of a wavefront sensor for coronagraphic aberrations \citep[e.g. ZELDA;][]{ndiaye2013}, coronagraphic phase diversity \citep[e.g. COFFEE;][]{paul2013,paul2014}, energy minimization or electric field conjugation \citep{borde2006,giveon2007}, or any combination of these methods. 

In conclusion, although the SLLC currently provides only a small gain in a relatively limited range of separation, any improvement of the low- to mid-frequencies (the easiest to correct) would immediately reflect on the performance in favor of the LSS mode with SLLC.

\section{First on-sky results}
\label{sec:first_on_sky_results}

\begin{figure}
  \centering
  \includegraphics[width=0.5\textwidth]{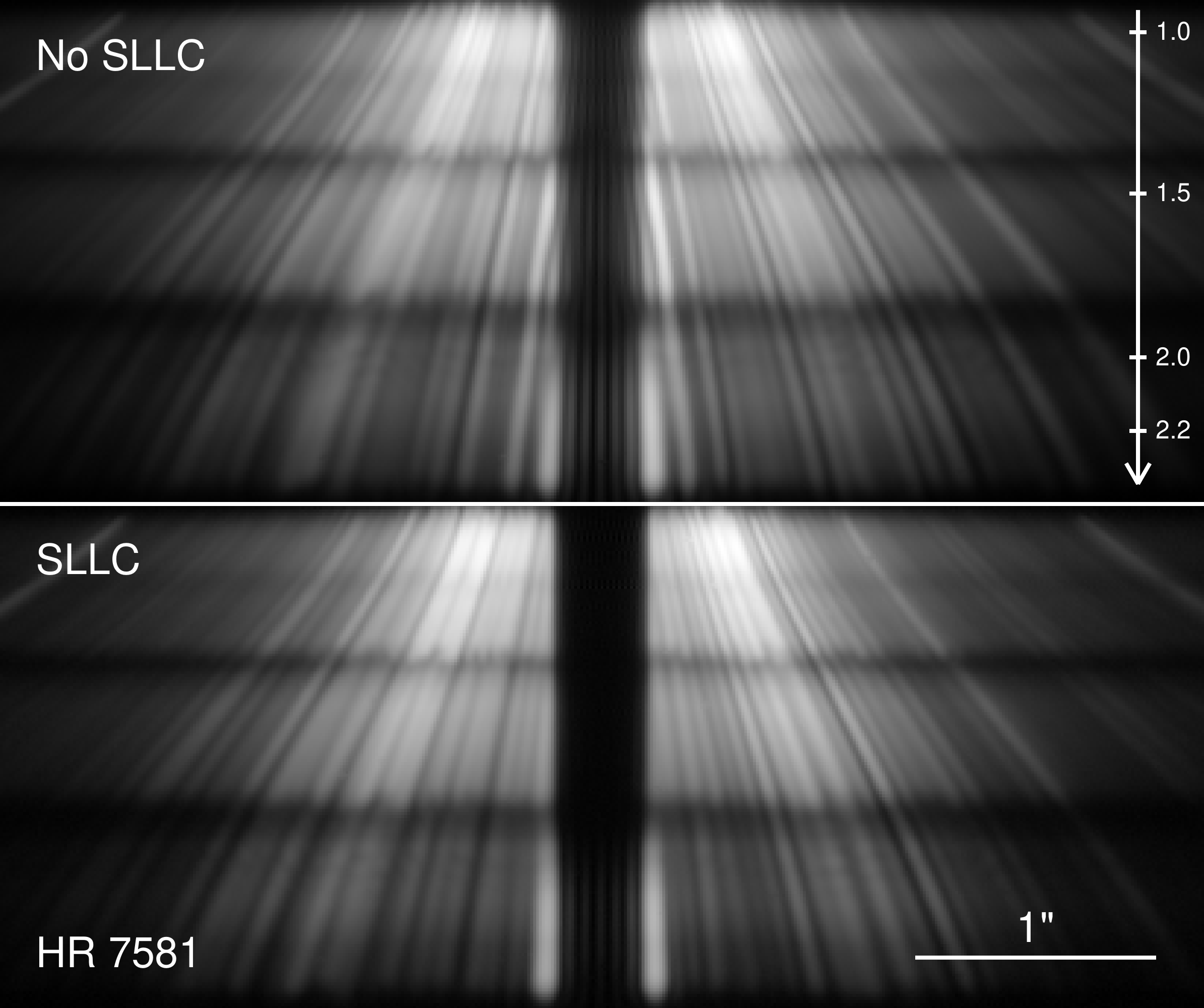}
  \caption{On-sky IRDIS/LSS data acquired on the star HR\,7581 without (top) and with the SLLC (bottom). The wavelength scale (in micron) is indicated with the downward white arrow. The total spectral range is 0.95--2.3~\mic, and the spatial extension is 2.5\as. The horizontal dark lines at $\sim$1.4~\mic and $\sim$1.9~\mic correspond to atmospheric absorption bands.}
  \label{fig:onsky_spectra}
\end{figure}

\begin{figure}
  \centering
  \includegraphics[width=0.5\textwidth]{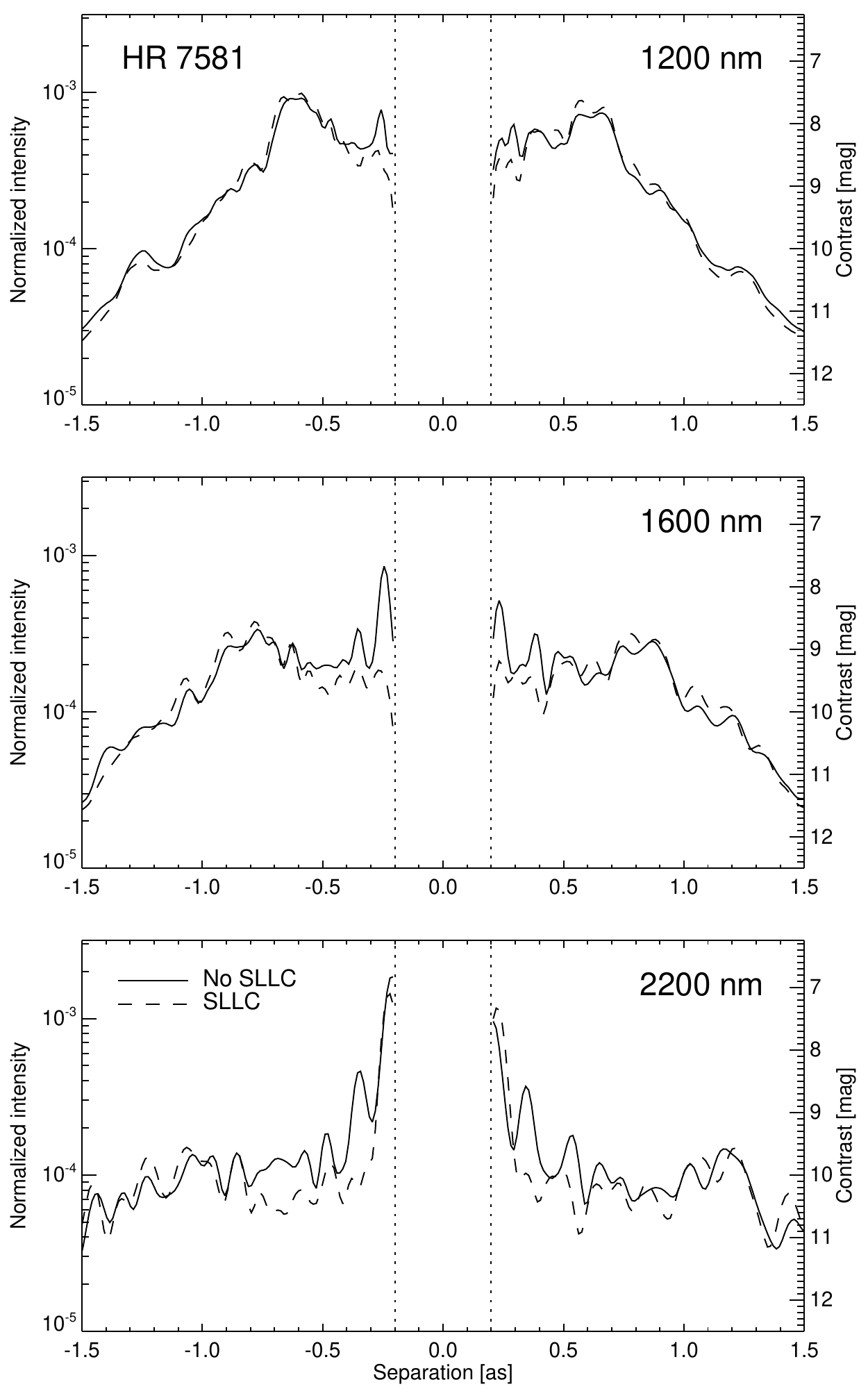}
  \caption{On-sky IRDIS/LSS coronagraphic intensity profiles with (dashed line) and without (plain line) the SLLC as a function of angular separation on both sides of the optical axis and normalized with respect to an off-axis PSF. Profiles are plotted at 1200~nm ($J$-band), 1600~nm ($H$-band), and 2200~nm ($K$-band).}
  \label{fig:onsky_spectro_plots}
\end{figure}

The internal measurements presented in the previous sections are encouraging, but they are somewhat disconnected from real-world observations where the amount of residual aberrations from the XAO system ($\sim$65~nm~rms) will dominate the overall error budget, and where the diffraction of the telescope central obscuration and spiders will certainly affect the performance. 

On the night of October 11, 2015 we had the opportunity of acquiring an on-sky data set with the SLLC for a first ``real-world'' assessment of the performance. We observed HR\,7581, a bright ($H=1.65$), K0 star between 00:00UT and 00:30UT. The observing conditions were average, with a reported DIMM seeing of 0.7--0.9\as, a coherence time slightly over 2~ms, and wind blowing from the west at a speed of $\sim$6~m/s. The target acquisition was done with the dedicated observing template for LSS observations. This template includes an automatic centering of the PSF behind the coronagraphic slit during the target acquisition. The centering appeared satisfactory and was not adjusted manually before the start of the data acquisition. We obtained two data sets, with and without the SLLC, each data set consisting of 4~min of deep coronagraphic data, 3~min of off-axis reference PSF, and an appropriate set of sky backgrounds. The wavelength calibration and flat fields were acquired in the morning as part of the daily calibrations of the instrument. The data were treated in exactly the same way as the internal spectroscopic data (see Sect.~\ref{sec:spectro_data_acquisition}). The spectra with and without SLLC are compared in Fig.~\ref{fig:onsky_spectra}, and coronagraphic profiles at three wavelengths are presented in Fig.~\ref{fig:onsky_spectro_plots}.

The non-SLLC and SLLC spectra look visually different at small separations (0.2--0.4\as). In this range, the non-SLLC spectrum is dominated by diffraction residuals similar to dispersed, bright, and dark Airy rings. This feature was already noted in the simulations of \citetalias{vigan2013} and was identified as the main limitation in the subsequent data analysis required for extracting the spectrum of planetary companions. These diffraction residuals are totally absent from the SLLC spectrum, which looks more or less similar to the data obtained on the internal source and showed in Fig. \ref{fig:data_simu_spectra} top frame. Figure~\ref{fig:onsky_spectro_plots} confirms the presence of the residuals in the non-SLLC data. Although not completely visible at 1200~nm, they become obvious at 1600~nm and 2200~nm. Even in $K$-band, where the mode is not fully optimized because of the size of the coronagraphic mask (see Sect.~\ref{sec:spectro_results}), there is a visible contrast gain of up to 1.5~mag in 0.5--0.7\as. In $H$-band, the gain is slightly smaller but still significant. These findings and the obtained contrasts are perfectly consistent with the results reported in \citetalias{vigan2013}. This is promising for future scientific results with the SLLC, because in \citetalias{vigan2013} we demonstrated that although the raw contrast gain with the SLLC may not be very impressive, the subsequent data analysis based on spectral differential imaging will yield better results when diffraction residuals are eliminated.

The results obtained on the internal source already showed some potential for the SLLC and highlighted the limitations of the current LSS mode. We now see that when considering the full telescope pupil, including both central obscuration and spiders, the difference between the SLLC and non-SLLC is even more accentuated. The influence of the spiders is expected to be noticeable at small inner-working angles with the SLLC because the apodization will broaden their diffraction pattern. In practice, observations should, if possible, be optimized so that the diffraction spikes of the spiders do not cross the slit during the time of the observation. In the case presented here, the orientation of the pupil was so that the diffraction spikes were far from the slit.

If the diffraction spikes of the spiders are not the primary limitation in our non-SLLC on-sky data, the only remaining possibility is the central obscuration of the telescope. As mentioned in Sect.~\ref{sec:sllc_perf_spectroscopy}, the Lyot stop in the LSS mode is a simple circular diaphragm with a size equal to 92\% of the pupil, which does not include any mask for the central obscuration. This is of course a severe limitation for the coronagraphic performance in LSS with the full telescope pupil. By contrast, the SLLC has been optimized \emph{\textup{by design}} to take the central obscuration into account (but not the spiders), which directly reflects on the on-sky performance.

These first on-sky results do not constitute an in-depth performance analysis of the apodized LSS mode of IRDIS, but they are nonetheless extremely encouraging for future scientific applications. We keep the analysis of the final scientific performance for a later
publication. This will take all the specific problems related to on-sky observations into account.

\section{Conclusions}
\label{sec:conclusions}

Characterizing directly-imaged giant planets through near-infrared spectroscopy is one of the main purposes of the new generation of high-contrast imagers and spectrographs to which SPHERE belongs. IRDIS, with its unique LSS mode that can reach resolutions up to 350 in YJH, is a particularly attractive solution, but the overall contrast performance is limited at very small angular separation because the coronagraph used in this mode is far from optimal. We have explored the possibility to improve this performance in \citetalias{vigan2013} by using the SLLC, an especially designed apodizer that basically suppresses the Airy rings above 4.53~\lsd without the need for a Lyot stop.

In this new paper, we have presented the specification, manufacturing, and testing of an SLLC prototype that has been installed inside SPHERE during the instrument reintegration at the Paranal observatory in 2014. While the prototype does not fully meet the specifications in terms of transmission, especially in the areas where very low transmission is required, the imaging data acquired with the SLLC show a gain of a factor 10 at 0.3\as and 5 at 0.5\as in raw contrast compared to standard imaging without the apodizer. In addition, using data acquired with an FPM, we have demonstrated that no Lyot stop is required to reach the full performance, which is a direct validation of the SLLC concept in imaging. Our simulations using phase and amplitude error maps measured inside the instrument show that in imaging we reach the limit set by the phase errors in the system ($\sim$35~nm~rms, without accounting for the DM dead or stuck actuators).

For LSS data the gain on the internal source is not as obvious when using the SLLC. In this mode, the FPM located at the center of the slit and the circular Lyot stop in the relayed pupil plane act as a form of CLC, which provides a partial attenuation of the diffraction even without using the SLLC. In the presence of the SLLC, there is a gain of raw contrast of only $\sim$1~mag in a limited range of separation (typically withing 0.5--0.6\as). However, our simulations show that the current LSS mode without SLLC is very close in performance to the absolute limit accessible in this mode. This means that if the performance of the system can be improved, for instance, by reducing the low- to mid-order aberrations using some form of wavefront control, the SLLC mode will  be significantly improved in performance while the non-SLLC mode will immediately reach its limit at very small angular separation.

The first on-sky data acquired with the SLLC are particularly interesting. They show that in presence of the telescope central obscuration, the gain of the SLLC is even more significant than on the internal source. Indeed, the Lyot stop in LSS does not include any mask for the central obscuration, which affects the coronagraphic performance in this mode. The result is visible in the data as a set of bright diffraction residuals at small angular separation. Thanks to its design, which takes a central obscuration into account, the SLLC is able to completely remove these residuals and provide a significant raw contrast gain in 0.2--0.7\as. This is perfectly consistent with the findings presented in \citetalias{vigan2013}. In addition, similarly to the results on the internal source, any reduction of the low- to mid-order phase aberrations will dramatically improve the performance of the LSS mode with SLLC with respect to the non-SLLC case. Of course, these preliminary results need to be confirmed and extended, but they constitute a very promising first step toward making the apodized LSS mode an official observing mode of SPHERE.

Finally, we conclude by stressing the importance of exploring simple high-contrast imaging solutions based solely on binary-shaped or gray apodizers, which require only a single pupil-plane device
instead of the complex coronagraphic setups that require in most cases two pupil planes and a focal plane for an apodizer, a Lyot stop, and an FPM, respectively. Indeed, all future extremely large telescopes (ELTs) include some exoplanetary science in their first-light science goals (see, e.g., \citet{eelt2009} for the European Extremely Large Telescope and \citet{skidmore2015} for the thirty-meter telescope). In particular, they all include high-resolution spectral characterization of giant exoplanets detected with previous generation instruments such as SPHERE or GPI. However, all the ELTs will start operations with general-purpose instruments, such as MICADO \citep{davies2010} and HARMONI \citep{thatte2014} for the E-ELT, which have not been specifically optimized for very high contrast. Specifically, these instruments will not rely on XAO, and may not even include fine tip-tilt control or atmospheric dispersion correctors, which are essential components when considering coronagraphs based on focal-plane occulters or phase masks \citep[e.g.,][]{mas2012}. In this context, exploring simple optical solutions that allow reaching high contrast at small inner-working angles without being sensitive to tip-tilt errors and/or relying on an XAO system is highly relevant. Theoretical solutions for complex apertures have started to be explored \citep[e.g.,][]{carlotti2013,carlotti2014,ndiaye2015a,ndiaye2015b}, but prototype testing on current generation instruments and/or test benches is also an essential first step. In this regard, the SLLC will hopefully be a precursor for the development of the first-light high-contrast instrumentation for ELTs.

\begin{acknowledgements}

The authors are very grateful to the referee, Wesley Traub, for reviewing this article and help in improving it. The authors would also like to thank C. Dorrer from Aktiwave LLC for his contribution on the manufacturing of the apodizer, and M. Kasper and G. Zins for their expert help with the SPHERE AO and control software. MN would like to thank R\'emi Soummer and Laurent Pueyo for their support. Finally, the authors are extremely grateful to Andreas Kaufer, Director of La Silla Paranal observatory, for authorizing them to acquire and present the very first on-sky SLLC data.

\smallskip \\

The SLLC prototypes were funded by the \emph{Action Incitative} program of \emph{Laboratoire d'Astrophysique de Marseille}. AV acknowledges support from the French National Research Agency (ANR) through the GUEPARD project grant ANR10-BLANC0504-01. This work is partially supported by the National Aeronautics and Space Administration under Grants NNX12AG05G and NNX14AD33G issued through the Astrophysics Research and Analysis (APRA) program (PI: R. Soummer). MN would like to acknowledge the ESO Chile Visiting Scientist program.

\smallskip \\

SPHERE is an instrument designed and built by a consortium consisting of IPAG (Grenoble, France), MPIA (Heidelberg, Germany), LAM (Marseille, France), LESIA (Paris, France), Laboratoire Lagrange (Nice, France), INAF - Osservatorio di Padova (Italy), Observatoire de Genève (Switzerland), ETH Zurich (Switzerland), NOVA (Netherlands), ONERA (France) and ASTRON (Netherlands) in collaboration with ESO. SPHERE was funded by ESO, with additional contributions from CNRS (France), MPIA (Germany), INAF (Italy), FINES (Switzerland) and NOVA (Netherlands). SPHERE also received funding from the European Commission Sixth and Seventh Framework Programmes as part of the Optical Infrared Coordination Network for Astronomy (OPTICON) under grant number RII3-Ct-2004-001566 for FP6 (2004-2008), grant number 226604 for FP7 (2009-2012) and grant number 312430 for FP7 (2013-2016).

\end{acknowledgements}

\bibliographystyle{aa}
\bibliography{paper}

\end{document}